\documentclass[final,3p,times,authoryear]{elsarticle}

\usepackage{amssymb}
\usepackage{amsmath}
\usepackage{svg}
\usepackage{url}
\usepackage{gensymb}
\usepackage{subfigure}
\usepackage{lineno}
\usepackage{todonotes}

\journal{International Journal of Human-Computer Studies}

\begin{document}

\begin{frontmatter}



\title{Generating Multimodal Textures with a Soft Hydro-Pneumatic Haptic Ring} 

 \author[1]{Ana Sanz Cozcolluela}
 \author[1]{Koen W\"osten}

 \affiliation[1]{organization={Department of Cognitive Robotics, Delft University of Technology},
addressline={Mekelweg 2},
city={Delft}, postcode={2628 CD},
country={The Netherlands}}

 \author[1]{Yasemin Vardar*}

\cortext[2]{Corresponding author}



\begin{abstract}
The growing adoption of extended reality (XR) has increased demand for wearable technologies that provide naturalistic tactile sensations while allowing users to interact freely with their environments using bare fingers. 
However, most existing wearable haptic devices support only a limited range of tactile modalities.
Here, we introduce a soft haptic ring and a data-driven rendering methodology for generating multimodal texture sensations. The device integrates pneumatic and hydraulic actuation to render roughness, thermal, and softness cues on the proximal phalanx. 
The ring can generate forces up to 1.75~N, produce displacements up to 0.27~mm within a 30–300 Hz operating range, and modulate display temperature by up to 25$^\circ$C within 65~s. The rendering methodology modulates these cues based on the user's exploratory actions: the hydraulic actuator conveys perceived temperature during static contact, while the pneumatic actuator generates pressure and vibration cues to convey softness and roughness during pressing and sliding gestures, respectively.
We evaluated the system in a user study with 15 participants who matched six virtual textures generated by the ring to their real counterparts and rated their perceived sensations using guided exploratory actions. Participants achieved an average texture-matching precision of 68\% and an $F_1$ score of 0.68. Adjective ratings confirmed that the ring produces distinct and perceptually rich stimuli across all rendered modalities. These findings demonstrate the potential of the proposed haptic ring and rendering methodology to deliver multimodal tactile cues away from the fingertip for immersive XR applications, enabling diverse tactile feedback while preserving natural physical interaction.

\end{abstract}




\end{frontmatter}



\section{Introduction}
Our tactile experiences with the surroundings profoundly shape our understanding of the world. Through touch, we interpret physical properties and manipulate objects with precision. Yet, in digital environments, these rich tactile interactions are mainly absent. Wearable haptic devices, which have attracted significant interest from academia and industry over the last decade, aim to bridge this gap by simulating naturalistic tactile cues using mechatronic components worn on the body. 

However, most state-of-the-art wearable haptic devices adopt glove-like designs, which, despite their intuitive interfaces, are often cumbersome, heavy, and restrictive~\citep{pacchioretti2017_wearable}. These designs can hinder hand movements and, since they typically cover the fingertips, pose challenges for finger tracking and natural interaction with the environment~\citep{Pacchierotti2016}. This issue is especially problematic in mixed and augmented reality applications, where users engage with physical environments enhanced by artificial sensations~\citep{teng2022_xr}.

Wearable haptic devices that relocate feedback from the fingertips to other regions of the hand offer a potential solution for embedding artificial tactile sensations in physical environments~\citep{Pacchierotti2016, riessenRelocatingThermalStimuli2024, Sun2022, deVlam2023_focused, sarac2025_relocation}.
By shifting stimulation away from the fingertip, these devices can mitigate inherent limitations of fingertip feedback, such as limited interaction area, interference with natural hand movements, and disruption of perception of surrounding objects, while still supporting dexterous tasks. Previous studies on distal haptic feedback have shown that users can perceive and discriminate tactile information, including texture, even when stimulation is applied away from the fingertips \citep{riessenRelocatingThermalStimuli2024, Gaudeni2019, Friesen2024, normand2025_augment}. These findings underscore the feasibility of relocation as a promising approach for wearable haptics.

However, despite their growing popularity, these devices face challenges related to weight, flexibility, and ease of wear. Many rely on rigid and bulky materials that conflict with the compliant nature of the human body, limiting both comfort and the fidelity of tactile interactions \citep{fleckWearableMultisensoryHaptic2025}. 

Recently, soft haptic devices, designed as rings~\citep{Talhan2020, Han2018}, wristbands~\citep{Zhang2021, Liu2021}, or sleeves~\citep{zhu2020_penusleeve, goetz2020_patch}, have been proposed to enhance wearability by conforming more naturally to the skin. Among these, pneumatic actuation has gained popularity for its fast response, cost-effectiveness, and high power-to-weight ratios. Nonetheless, hydraulic actuation offers an alternative with improved drive efficiency and dynamic performance. 

These soft wearable devices can also deliver a range of tactile sensations, enhancing the naturalism and information richness of virtual interactions. For instance, the multimodal soft ring by \cite{Talhan2020} provided static pressure, high-frequency vibration, and impact force feedback through controlled pneumatic inflation. Similarly, the hydraulic thimble demonstrated by \cite{Han2018} combined pressure and vibration cues with thermal stimulation by circulating water at different temperatures. Other studies, such as \cite{Zhang2021, Liu2021, goetz2020_patch}, integrated pressure and thermal actuation in soft wristbands or sleeves. Table~\ref{tab:table1} compares the technical specifications of state-of-the-art soft wearable haptic devices capable of delivering multimodal tactile stimuli. While our primary focus is on designs that provide relocated tactile feedback, the table also includes thimble interfaces that directly actuate the fingertip, serving as a reference for recent developments.

\begin{table*}[!t]
\centering
\scriptsize
\caption{Comparison of state-of-the-art soft wearable haptic devices capable of delivering multimodal tactile feedback---pressure (P), vibration (V), and/or temperature (T)--- to either at the fingertip or at a relocated body location, including the proposed haptic ring. ``Pneu'', ``Hydr'', ``EHydr'', and ``Therm'' denote pneumatic, hydraulic, electrohydraulic, and thermal actuation mechanisms, respectively. ``h'' and ``c'' indicate heating and cooling capabilities. The horizontal line separates relocated designs from the ones with fingertip-based feedback. Glove-based systems are excluded.\\}
\begin{tabular}{ c c c c c c c c c} 
 \textbf{Reference} & \textbf{Interface} & \textbf{Actuation} & \textbf{Feedback} & \textbf{Max Force} & \textbf{Max Vib. Frequency} & \textbf{Weight} & \textbf{Temp. Range}\\ [0.5ex] 
 \hline
 \cite{Han2018} & Thimble & Hydr. & P + V + T (h+c) & 2.6~N & 100~Hz & - & 15-40~°C\\
\cite{Talhan2022} & Thimble &  Pneu.  & P + V & 7 N  & 250~Hz & 8~g &-\\
\cite{Hashem2022} & Thimble & Pneu.  & P + V & 5.5 N & 100~Hz & -&-\\
\cite{purnendu2023_electrohydraulic} & Thimble & EHydr. & P + V & - & 700~Hz & $\approx$2~g  & -\\
\cite{Shao2025_electrohydraulic} & Thimble & EHydr. & P + V & 8~N & 500~Hz &$\approx$8~g & -\\
\cite{Wang2025_ttouch} & Thimble & EHydr + Therm. & P + V + T (h+c) & 0.47~N& 10~Hz & $\approx$8~g&  20-40 °C\\
\cite{Chen2026} & Thimble &  Pneu. + Therm. & P + T (h) & 8.93 N  & - & 2 g & 25-50 °C\\
\hline
\cite{zhu2020_penusleeve} & Sleeve & Pneu. & P + V & 4~N & 50~Hz & 26~g& -\\
\cite{goetz2020_patch} & Sleeve & Hydr. & P + T & - & 0.22~Hz &- & 17-42~°C\\
\cite{Zhang2021} & Wristband & Pneu. + Therm. & P + T (h+c) & 10 N  & - &- & 15-40~°C\\
\cite{Liu2021} & Wristband &  Pneu. + Hydr. & P + T (h+c) & - & - & -& 25-42~°C\\
\cite{Talhan2020} & Ring & Pneu. & P + V & 6.3 N & 250 Hz  & 4.5 g &-\\
 \textbf{Current work} & Ring & Pneu. + Hydr. & P + V + T (h+c) & 1.75 N & 300 Hz & $\approx9$g & 10-40~°C
\end{tabular}
    \label{tab:table1}
\end{table*}

Despite their potential for displaying a wide range of tactile sensations, most wearable haptic devices, both soft and rigid, primarily provide basic stimuli, such as discrete forces, temperature changes, or vibrations, rather than rich textural experiences. Few studies explicitly address tactile texture rendering. For instance, multimodal haptic thimbles by \cite{Talhan2022, Hashem2022} or the vibrotactile rings by \cite{Gaudeni2019, Friesen2024, normand2025_augment} primarily encode roughness through vibration, but fall short in conveying other key perceptual dimensions, such as softness and temperature ~\citep{Okamoto2013}. Supporting these additional cues is therefore important for achieving richer and more realistic tactile experiences in digital environments. 

Moreover, no existing rendering methodology fully conveys the multimodal sensations of real surfaces. Current approaches primarily focus on simplified artificial textures, thus failing to capture the full sensory richness of real materials. For example, recent research on non-wearable multimodal haptic devices, such as tool-based \citep{Kodak2023} or surface displays \citep{vardar2025_multimodal}, has demonstrated the presentation of multiple sensory modalities for artificial textures. However, these approaches rely on stimuli defined by a limited set of parameters constrained by actuator capabilities, rather than signals derived from interactions with real materials. To the best of our knowledge, no wearable haptic device or rendering methodology has been proposed to convey the multimodal tactile properties of real materials, nor has prior work examined how such cues are perceived when presented at relocated locations on the hand. 

\begin{figure}[t]
    \centering
    \includegraphics[width=0.6\linewidth]{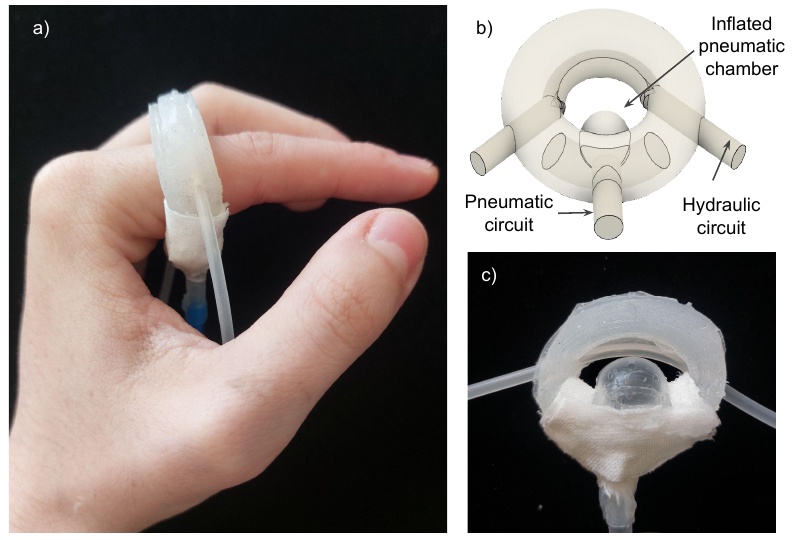}
    \caption{a) The soft haptic ring worn on the finger. b) The ring can provide multisensory cutaneous feedback via fluidic actuators. A pneumatic chamber at the finger's base provides pressure and vibration cues, while a hydraulic tube routed in a half-loop around the proximal phalanx transmits thermal sensations. c) The pneumatic actuator in inflated mode.}
    \label{fig:Ring}
\end{figure}

In this study, we introduce a soft haptic ring and a data-driven rendering approach for displaying multimodal textures (Figure~\ref{fig:Ring}a). The ring, fabricated entirely from silicone, can deliver vibrotactile, pressure, and thermal cues to the proximal phalanx while allowing full-hand motion and supporting natural exploratory procedures used in tactile perception. The design integrates a hydraulic circuit for thermal stimulation with a pneumatic chamber that produces pressure and vibrotactile feedback. The rendering methodology selects modality-specific signals based on the user’s exploratory actions and the material being rendered: pressure cues convey softness, vibrotactile cues encode dynamic texture, and thermal cues provide contextual information about the material’s overall warmth or coolness.

We first characterized the device to assess its mechanical performance. Then we validated both the design and rendering methodology through a user study in which 15 participants matched six virtual textures generated by the ring to their real counterparts and rated their perceived sensations. To the best of our knowledge, this work presents the first soft ring interface that actuates the proximal phalanx of the finger while integrating multiple actuation mechanisms and three haptic modalities within a single device. Furthermore, no existing rendering methodology combines these modalities within a unified, data-driven approach. This work also provides the first investigation into how multimodal textures rendered at a relocated location by such a device can be identified and perceived.

This paper is organized as follows. Section~\ref{sec:device} presents the design of the soft, multimodal haptic ring, describes its fabrication process, and details the device control and mechanical characterization. Section~\ref{sec:rendering} explains the texture rendering methodology. Section~\ref{sec:experiment} outlines the user study conducted to validate the proposed device and rendering approach, and presents the corresponding results. Section~\ref{sec:discussion} discusses the study and gives recommendations for future research. 

\section{Soft Multimodal Haptic Ring}\label{sec:device}

\subsection{Design and Fabrication}\label{sec:design}

We aimed to overcome the limitations of existing works by developing a new device that satisfies the following key requirements:

\begin{enumerate}
\item Provide a flexible, lightweight, and compact form factor for comfortable wear.
\item Leave the user's fingertip unobstructed to enable natural interaction with the physical environment. 
\item Deliver multimodal tactile stimuli (pressure, thermal, and vibration) relevant for texture perception. 
\end{enumerate}

We addressed the first and second requirements by designing a soft ring fabricated entirely from silicone as the haptic interface. The ring design ensures that the fingertip remains unobstructed, allowing for natural hand movements and unrestricted primary exploration procedures, such as pressing, sliding, and static contact~\citep{lederman1987_hand}. The choice of a ring design over a bracelet was also motivated by a prior study by \citet{Gaudeni2019}, which demonstrated that frequency discrimination thresholds for vibrotactile stimuli are significantly higher at the wrist than at the fingertip or distal phalanx, which exhibited comparable sensitivity levels. Moreover, the use of silicone maximizes flexibility, lightness, and compactness, while also facilitating easy fabrication of different sizes that accommodate various finger dimensions.

The third requirement presents a challenge when working with soft interfaces. Thermal stimuli are typically generated using thermoelectric devices~\citep{Jones2008}, but their rigid nature can jeopardize the flexibility of the soft ring. Similarly, although piezoelectric actuators can produce complex vibration patterns~\citep{biswas2019_emerging}, their integration would compromise the soft character of the ring. To address these challenges, we adopted a hybrid actuation design that combines hydraulic and pneumatic systems (Figure~\ref{fig:Ring}b). The hydraulic system circulates water at controlled temperatures through the ring to provide the desired thermal stimuli. Independent of the thermal feedback, the pneumatic system inflates a chamber within the ring to create vibratory cues through rapid inflation and deflation. Alternatively, by slowly and steadily inflating the chamber, the pneumatic system can provide pressure stimuli. This dual-actuation approach enables the delivery of thermal stimuli either independently or in combination with either pressure or vibration cues.

While complete independence and simultaneous rendering of all modalities remain desirable, we chose to use a single actuator for both vibration and pressure to maintain a sleek and compact design. The coordinated delivery of these cues for reproducing texture sensations is managed by the rendering algorithm described in Section~\ref{sec:rendering}.

In our design, the pneumatic chamber is positioned on the underside of the ring, in contact with the ventral side of the finger's proximal phalanx (Figure~\ref{fig:Ring}b and c). The ventral side of the finger is covered with glabrous skin, which has been shown to exhibit greater sensitivity to vibrotactile stimuli due to the high density of fast adapting mechanoreceptors \citep{mahns2006_hairy}. On the upper side of the ring, in contact with the dorsal phalanx, a thin silicone tube forming part of the hydraulic circuit is placed. Instead of embedding a hydraulic chamber, we inserted a silicone tube directly into the ring to maximize thermal conductivity. The dorsal side of the finger was selected for thermal stimulation, as prior research has shown that nonglabrous (hairy) skin exhibits higher thermal sensitivity \citep{wakolbinger2014, filingeriThermosensoryMicromappingWarm2018}. 

For the fabrication of our soft ring, we chose a highly stretchable silicone rubber (EcoFlex 00-30, Smooth-On), a material whose softness, strength, and stretchability make it an ideal choice for wearable and skin-attachable devices~\citep{Rameshwar2023}. Additionally, its elastic modulus (100 to 125~kPa) is comparable to that of human skin \citep{Kalra2016MechanicalBO}, enabling a comfortable and compliant interface. 

The ring fabrication process involved multiple steps. First, we 3D-printed two identical plastic molds to form the two halves of the ring. These molds, made from polylactic acid (PLA), were designed to split the ring along its transversal plane and were further divided into two parts to facilitate demolding. We prepared three mold variations to accommodate different finger sizes with internal diameters of 15, 18.5, and 22~mm. We ensured that the pneumatic chamber maintained a constant volume across all ring sizes to exhibit similar inflation characteristics. 

The next step in the fabrication process was casting the molds. This step begins with mixing the parts of EcoFlex silicone rubber (Part A and Part B). After thorough stirring, the molds were coated with a release agent to facilitate demolding, and the mixture was poured in and then left to cure for at least four hours at room temperature. Once cured, the two halves were bonded using the same silicone mix. At this stage, a 4~mm diameter silicone tube was inserted between the halves, allowing for future inflation of the pneumatic chamber. Precise alignment of all components was crucial to ensure a durable and well-functioning final product. We also reinforced the insertion area by lining it with polytetrafluoroethylene (PTFE) tape, coating with a stiffer layer of silicone rubber (DragonSkin 30, Smooth-On), and curing for seventeen hours. The higher Young's modulus of this silicone mix, combined with the thin PTFE tape layer, provided structural support by preventing leakage or bursting at high pressures. During this process, it was crucial to apply the new silicone mix only at the tube insertion while ensuring the inner part of the ring remained uncovered to preserve the actuator's inflation characteristics.

During the final step of the manufacturing process, a strain-limiting fabric layer was applied to the ring's upper, lower, and outer surfaces using  silicone adhesive (Sil-Poxy, Smooth-On). This process ensured that when air filled the pneumatic chamber, only the ring's inner surface would inflate and provide localized pressures at the frontal side of the phalanx. Additionally, during this step, a thin 2~mm silicone tube connected to the hydraulic circuit was embedded on the inner surface of the ring to present thermal cues.

\subsection{Control}\label{sec:control}

\begin{figure*}[b]
    \centering
    \includegraphics[width=\textwidth]{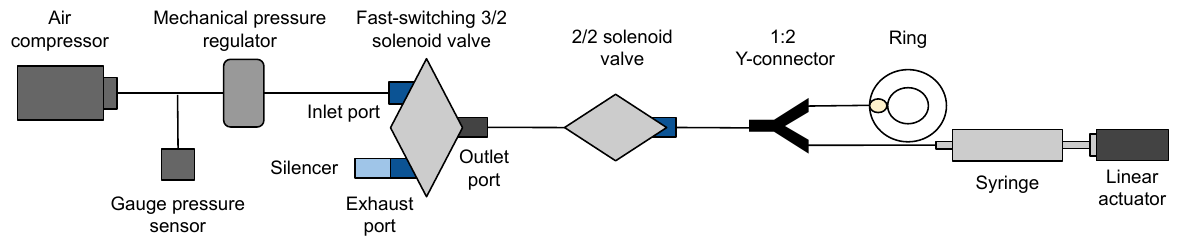}
    \caption{Schematic of the pneumatic circuit. The 2/2 solenoid valve remains open to generate vibration cues and is closed for pressure cues. }
    \label{fig:PneumaticCircuit}
\end{figure*}

In our soft haptic ring, the pressure and vibration cues are generated through a pneumatic circuit connected to the ring's pneumatic chamber, as seen in Figure~\ref{fig:PneumaticCircuit}. This circuit includes two different sections connected in parallel to generate pressure and vibration stimuli, respectively.

The primary component for generating vibration cues is a fast-switching three-way solenoid pneumatic valve (MHE2-MS1H-3/2G-M7, Festo), selected for its high 300~Hz switching frequency. The valve's inlet port is supplied by an air compressor (FD-186, Fengda), with the pressure controlled at 75~kPa using a mechanical regulator (AR-200, LNCN) and monitored by a digital pressure gauge. This pressure level ensures perceivable inflation of the actuator while preventing excessive force that could lead to rupture. 

In the unlikely event of a burst, the energy released would be minimal ($<$ 3~J), even at 75~kPa above atmospheric pressure, due to the small air volume contained in the system ($\approx$25~ml). Moreover, the absence of rigid components further mitigates potential risk. As an additional safety measure, the lower limit of the valve was set at 30~Hz, as low-frequency vibrations (1-20~Hz) could cause the silicone ring to balloon and burst.  

The valve's output port is connected to the ring via a 1:2 Y-connector, while the exhaust port is fitted with a silencer to reduce noise. This configuration allows air to flow freely into the ring during active valve state, inflating the actuator, and to vent rapidly through the exhaust when the valve is inactive, enabling rapid deflation. Consequently, the quick activation and deactivation of the solenoid valve produces vibratory cues through alternating inflation and deflation of the ring.

Although this valve could theoretically provide pressure cues by adjusting switching times, it proved challenging to control inflation rates precisely using this method. Additionally, this setup did not allow air to be retained inside the ring or expelled at variable speeds, limiting its ability to simulate materials of different stiffness or maintain inflation for prolonged contact sensations. Hence, a pressure rendering system was added in parallel to the vibration valve via the Y-connector to address these limitations. This pressure rendering system consists of a linear actuator controlling a syringe connected to the ring. The ring can exhibit various inflation profiles by adjusting the actuator's speed. However, due to the low operating pressures of the circuit, the fast-switching solenoid valve exhibited minor leakage in its closed state, preventing the ring from maintaining inflation. Hence, an additional normally closed two-way solenoid valve (MHA1-M1H-2/2G-0.9-PI, Festo) was inserted between the vibration valve and the Y-connector. In this way, the valve remains open to provide vibratory stimuli, allowing air to flow to the ring. Conversely, when providing pressure stimuli, the valve closes, enabling the ring to inflate, maintain its inflation, or deflate based on the operation of the linear actuator controlling the syringe. The valve thus determines the rendering mode, either vibration or pressure, preventing both modalities from operating at the same time.

\begin{figure}[t]
    \centering
    \includegraphics[width=0.65\linewidth]{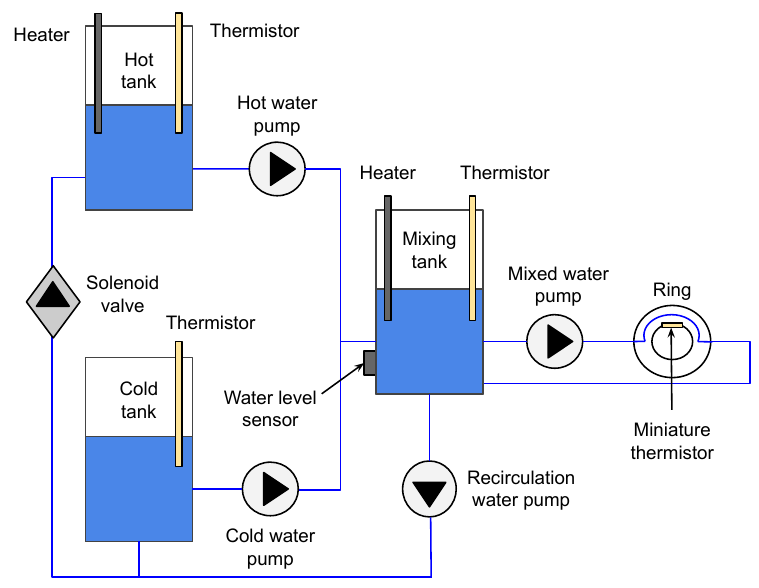}
    \caption{\centering{Schematic of the hydraulic circuit.}}
    \label{fig:HydraulicCircuit}
\end{figure}

Thermal cues were delivered by regulating the water temperature inside the silicone tube via a hydraulic circuit, as illustrated in Figure~\ref{fig:HydraulicCircuit}. The circuit comprised a hot and cold tank, whose water was mixed in a central container before being pumped to the ring. The hot water tank maintained a constant temperature of 42.5~°C, regulated by a 300~W submersible heating element (860-6921, RS PRO). The cold water tank was kept below 5~°C by continuously adding ice. Water from these tanks was transferred to the mixing container using two independent pumps (KW-1646, Kiwi Electronics), from which a third pump directed the mixed water through the silicone tube embedded in the ring. After circulating, the water returned to the mixing tank, completing the cycle.

Although using a mixing tank rather than directly pumping hot or cold water to the ring introduced a slight delay in reaching target temperatures, it ensured more uniform mixing and better temperature stability. To minimize delays, the water level in the mixing tank was kept at a minimum by using a non-contact liquid level sensor (XKC-Y25-NPN), which triggered a hydraulic pump to return excess water to the hot and cold tanks. A solenoid valve (ADA-997, Adafruit) prevented backflow and unintended mixing between tanks. Additionally, a 100~W submersible heating element (860-7163, RS PRO) in the mixing tank activated when the temperature dropped below a set threshold, further reducing waiting times.

Water temperatures in the hot, cold, and mixing tanks were continuously monitored using submersible NTC thermistors (B57861S0103F040, EPCOS). To ensure accurate user-experienced temperature control, a miniature NTC thermistor (GA10K3MBD1, TE Connectivity) was positioned between the silicone tube and the ring, measuring the external surface temperature of the hydraulic tube.

The entire system is controlled by a microcontroller (Mega, Arduino Inc.). Dual H-bridge motor drivers (L298N, Kiwi Electronics) power and independently control the hydraulic pumps, linear actuator, and pneumatic valves. Meanwhile, each heating element is regulated by a relay module, ensuring safe and independent operation. While the pneumatic components are managed by directly toggling their activation states, the hydraulic system is regulated by a proportional controller. This controller dynamically adjusts pump speeds based on the difference between the target display temperature and the measured value to ensure precise temperature control. The specifics of this controller are elaborated in \ref{app:thermal-control-logic}.

\subsection{Characterization}\label{sec:characterization}
We characterized the ring to assess its mechanical and thermal performance, verify the functionality of each actuation modality, and inform the development of the rendering methodology. Each modality was evaluated independently, with all other actuation modes disabled to avoid cross-interference. 

\subsubsection*{Pressure:}

\begin{figure*}[t!]
    \centering
    \includegraphics[width=\textwidth]{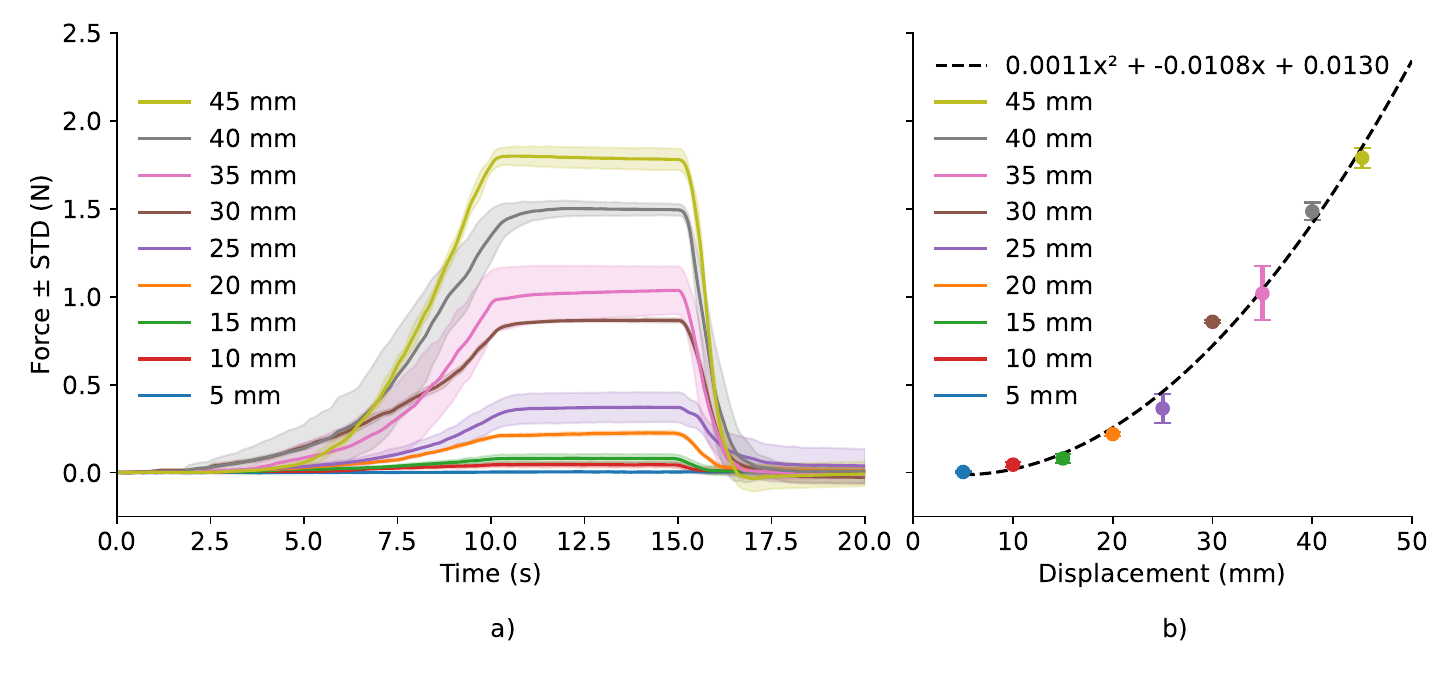}
    \caption{Results of the characterization experiments for the pressure modality of the ring. a) Normal force measured during inflation and deflation at different linear actuator displacements as a function of time. Color-coded lines and shaded regions indicate mean and standard deviation across five repeated trials for each condition. b) Corresponding average force values during steady-state phase with the fitted curve, demonstrating a  quadratic relationship ($R^2 = 0.989$) between linear actuator displacement and force.}
    \label{fig:forces}
\end{figure*}

The ring’s capability to generate pressure stimuli---produced through the slow inflation and deflation of the pneumatic actuator---was evaluated by measuring the output force as a function of the linear actuator’s displacement. For this purpose, a force sensor (FSG020WNPB, Honeywell) was encased in a saddle-like jig and mounted inside the ring (Figure~\ref{fig:force_setup}). During measurements, the ring was suspended by its pneumatic tubing, allowing it to hang freely to prevent external mechanical interference (see Figure~\ref{fig:force_thermal_setup}). The sensor was oriented upward to measure normal force exerted by the inflating silicone chamber. Inflation was achieved by driving a linear actuator that moved the plunger of a syringe, thereby compressing the air within the pneumatic circuit. The relationship between plunger displacement and output normal force was measured to quantify the mechanical output range. Data was recorded using a acquisition board (PCIe-6323, National Instruments) connected through an SCB-68A terminal block. Each measurement cycle consisted of a 10-second transient phase to reach the target linear actuator extension, followed by a 5-second steady state and subsequent deflation phase. Measurements were repeated five times for nine different extensions. Between runs, the saddle jig was repositioned to release residual stress caused by the silicone gripping the jig during deflation. Figure~\ref{fig:forces}a shows the measured force over time, with error bars indicating the standard deviation. As shown here, the pneumatic actuator can generate forces up to 1.75~N with a linear stroke of 45~mm. Figure~\ref{fig:forces}b presents the corresponding average force values during the steady-state phase, with error bars indicating standard deviation within that phase. The results reveal a strong quadratic relationship between the linear actuator displacement and measured force ($R^2 = 0.989$). 

\subsubsection*{Thermal:}
The ability of the ring to generate the desired thermal stimuli was evaluated by measuring the temperature changes on the silicone tube that transported water through the ring, with the temperature sensor placed using a single winding of polytetrafluoroethylene (PTFE) tape. During testing, the ring was suspended from the pneumatic tube, allowing it to hang freely in the air to avoid heat transfer to external surfaces (Figure~\ref{fig:force_thermal_setup}). Two sets of measurements were conducted: one starting at 25$^\circ$C, near room temperature, to capture the system's startup behavior, and another starting at 35$^\circ$C, representing the average initial temperature during experiments. Before each run, both the mixing tanks and ring were allowed to reach thermal steady state. The cold tank was maintained below 5$^\circ$C and the hot tank at 42.5$^\circ$C, with reheating or cooling performed as necessary between runs. During each measurement, the temperature was held constant for the first 5~s, after which the target temperature was set to one of 10, 15, 20, 25, 30, 35, or 40$^\circ$C and maintained until a 90~s limit was reached. Each measurement was repeated twice. 

\begin{figure*}[t!]
    \centering
    \includegraphics[width=0.97\textwidth]{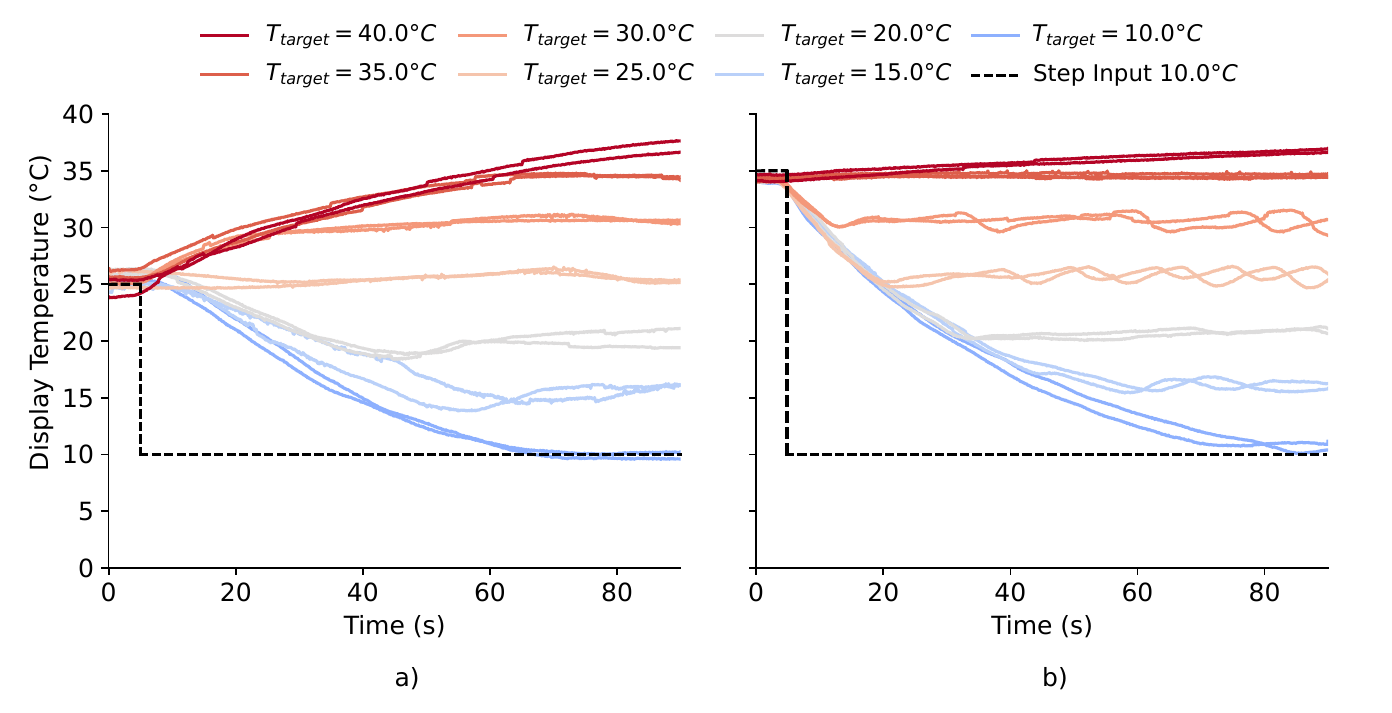}
    \caption{Results of the characterization experiments for the thermal modality of the ring. a) Temperature of the silicone tube over time, starting at 25$^\circ$C, illustrating the system's behavior during initial startup. b) Temperature response starting at 35$^\circ$C, representing typical operating condition.}
    \label{fig:temperature}
\end{figure*}

Figure~\ref{fig:temperature} presents the resulting time-temperature responses as the ring transitions to assigned temperature setpoints. After the initial 5~s hold, each trace departs from its starting value and converges toward the target, exhibiting the oscillatory behavior expected from a proportional controller regulating the hot-cold mixing. As shown, the actuator can modify display temperature by up to 25$^\circ$C within 65~s, with shorter response times for smaller temperature changes. Although larger overshoots would typically be expected for larger temperature steps, this effect is attenuated in our system because the mixed water is continuously recirculated. As the display temperature shifts, the effective temperature difference between the hot and cold tanks decreases, thereby limiting the magnitude of overshoot across conditions.

\subsubsection*{Vibration:}
The ring's ability to generate vibration stimuli was characterized by measuring the pneumatic actuator's response at different valve frequencies. The ring does not directly control vibration amplitude; instead, the displacement arises from the system's mechanical response to the applied pressure at the excitation frequency. Therefore, the values reported in this section represent the measured steady-state behavior of the ring under the specified operating conditions.

Actuator displacements were captured using a high-speed camera (VEO-E 340L, Phantom) at a resolution of 2560$\times$200 and 3000~fps. The camera was equipped with a macro lens (150~mm f/2.8, Irix) and the body was positioned at 37.5~cm from the ring. The setup can be found in Figure~\ref{fig:displacement_setup}. In this setup, 86.5~pixels correspond to 1~mm at the plane of intersection with the ring.

Each measurement consisted of a single-frequency burst lasting 4~s, including a ramp-up phase of less than 1~s, providing a 3~s Fast Fourier Transform (FFT) analysis window. Excitation frequencies were chosen as a geometric progression of 20 points spanning 30-300~Hz. The system was driven with a binary on/off signal applied to the fast-switching solenoid valve at the excitation frequency, with all measurements conducted at a constant input pressure of 75~kPa above atmospheric. To minimize interference from ambient light, the setup was covered with a black cloth during recordings. The displacement at each excitation frequency was obtained from the FFT of the steady-state segment, as detailed in \ref{app3}. The amplitudes were corrected for half-spectrum scaling and sample size. This procedure was repeated for all tested frequencies to construct the complete displacement magnitude response curve. 
The single-sample measurement noise was estimated from zero-displacement recordings, yielding a standard deviation of $0.0023$~mm 
and empirical measurements for the 9000-sample analysis window show a noise floor of $0.000114$~mm.

The vibration measurements were obtained under unloaded conditions, without skin contact. Although the vibration magnitude is expected to decrease when the device is mounted on a finger, the relative frequency-dependent behavior in unloaded conditions remains informative because the actuator's input energy, determined by the supply pressure and gated by the on/off valve, remains constant between loaded and unloaded conditions. 

Additionally, to assess potential material degradation resulting from the continuous mechanical stress of repeated stretching and relaxation, which may reduce the ring’s elasticity over time, we compared displacement measurements from a ring that had been extensively used over a one-year period with those from a newly fabricated ring, both produced using identical manufacturing procedures.

\begin{figure}[t]
    \centering
    \includegraphics[width=\linewidth]{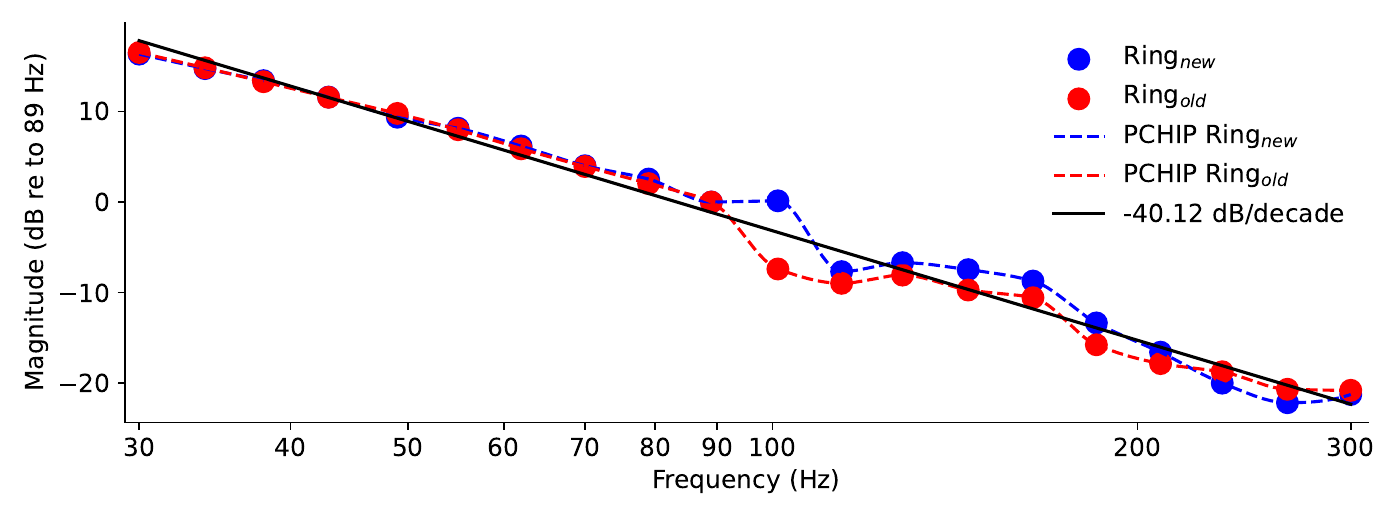}
    \caption{Results of the characterization experiments for the vibration modality of the ring. The magnitude response of the pneumatic actuator's displacement for the old (red) and new (blue) rings as a function of excitation frequency, measured under a constant input air pressure of 75~kPa above atmospheric. Each measurement point is shown as a large filled circle; color-coded dashed lines represent the fitted Piecewise Cubic Hermite Interpolating Polynomial (PCHIP) curves, and the solid black line indicates the linear fit. }
    \label{fig:magnitude_response}
\end{figure}

Figure~\ref{fig:magnitude_response} shows the variation in displacement amplitude with frequency under the fixed input drive of 75~kPa, relative to the sample closest to the geometric mean of the tested bandwidth ($\sqrt{30 \cdot 300} \approx 94.87$); in this case, the 89~Hz sample. At this frequency, the vibration magnitude was 0.03~mm for the new ring and 0.04~mm for the old ring. The resulting displacement data were then interpolated using a Piecewise Cubic Hermite Interpolating Polynomial (PCHIP), which preserves the overall shape of the signal. The system (both with old and new rings) attenuated higher-frequency inputs smoothly, with no clear resonance peaks. The consistent monotonic decay of 40.12~dB/dec suggests no oscillatory or unstable modes within the 30–300~Hz band. Overall, across the full 30–300~Hz operating range, the measured steady-state displacement of the ring surface decreased from approximately 0.27~mm at 30~Hz to about 0.003~mm at 300~Hz. 

\section{Action-Based Texture Rendering}\label{sec:rendering}
\begin{figure}[t]
    \centering
    \includegraphics[width=0.59\linewidth]{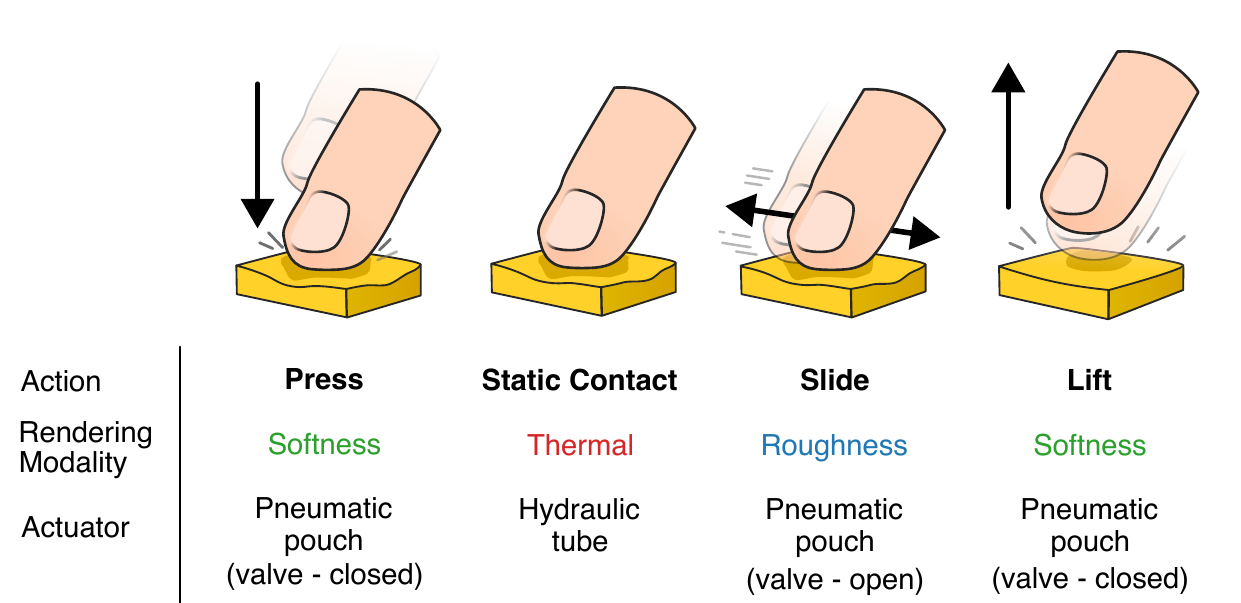}
    \caption{Rendering modalities and actuation mechanisms of the soft haptic ring in the proposed action-based rendering methodology.}
    \label{fig:actions}
\end{figure}

To complement the device design and fully exploit its multimodal capabilities, we developed a rendering strategy that determines when and how each tactile modality should be activated during interaction with virtual textures. This method dynamically delivers the most perceptually relevant tactile cues based on the users' exploratory actions. Prior research by \cite{lederman1987_hand} identified distinct exploratory strategies used to perceive different tactile properties---for example sliding for surface roughness, pressing for softness, and static contact for thermal cues. Our rendering methodology leverages these natural interaction strategies to selectively trigger each modality while accounting for the capabilities and limitations of the proposed ring device, thereby compensating for the shared actuation of pressure and vibration cues and mitigating delays inherent in the thermal system (Figure~\ref{fig:actions}). 

With this approach, we aimed to replicate the experience of exploring surfaces with bare fingers by actuating the proximal phalanx through the ring device. To inform the rendering of these tactile cues, we utilized the SENS3 database by \cite{Balasubramanian2024}, which provides multimodal tactile interaction data collected from participants exploring fifty surfaces. The dataset includes high-resolution top-view surface images, auditory recordings of interaction sounds, and detailed measurements of finger-surface interactions that capture the natural perceptual strategies humans use when assessing textures, such as static contact, pressing, tapping, and sliding.

Although the SENS3 database includes high-fidelity finger-surface interaction recordings, the physical constraints of our fluidic actuators made direct reproduction of these signals infeasible. Instead, our objective was to synthesize a range of tactile sensations that evoke the perceptual qualities of real textures rather than achieve strict one-to-one signal matching. Below, we describe how we processed the SENS3 recordings to render sensations of softness, temperature, and roughness, and we outline the specific device-level limitations that guided the rendering design for each modality.

It is important to note that in this approach, the rendered surface properties originate solely from recorded interaction profiles, which capture fingertip deformation or temperature changes during contact with real surfaces, but are applied to the user’s proximal phalanx while the finger moves in mid-air. In our implementation, a graphical user interface instructs the user to perform specific exploratory actions (e.g., sliding, pressing, or static contact) and constrains the motion to predetermined trajectories. The method does not incorporate the real-time forces or motion that would occur if the ring-wearing user freely explored a physical surface (e.g., a flat material), nor does it account for the resulting deformation, temperature changes, or perceptual interactions at the fingertip. Such capabilities would be necessary to overlay virtual textures onto real materials, as required in mixed-reality applications.

\subsubsection*{Softness Rendering (Press \& Lift Action):}
The softness of an object is typically associated with perceived resistive and deformation cues, producing both kinesthetic and cutaneous sensations when force is applied~\citep{Srinivasan1995TactualDO}. Traditionally, object softness has been rendered using force-feedback devices, which dynamically modulate the displayed force ($F$) based on the depth of indentation ($d$) and the desired spring constant ($k$), following the linear relationship of $F = kd$ \citep{srinavasan1997_haptics}. Recent works \citep{scilingo2010_softness, mete2024_sori} have also proposed new haptic devices that can modulate resistance and deformation cues by independently controlling the stiffness and finger contact area.

Our device renders softness by applying pressure cues through the embedded pneumatic pouch driven by a linear actuator. The applied air pressure modulates both resistance and deformation cues on the skin as the balloon-like pouch expands and contracts in response to pressure changes. Consequently, material softness can be rendered by controlling the linear actuator displacement. 

\begin{figure*}[!t]
    \centering\includegraphics[width=\textwidth]{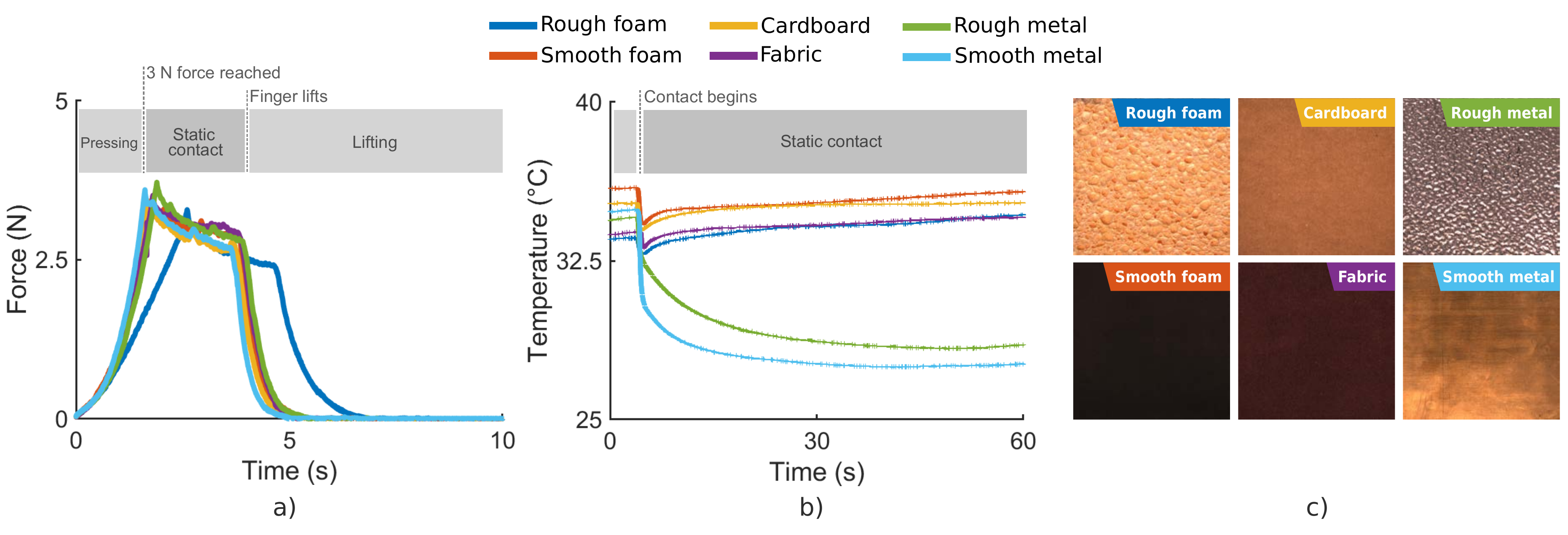} 
    \caption{The texture data used from the SENS3 Database~\citep{Balasubramanian2024}: rough metal (Metal 1), smooth metal (Metal 5), rough foam (Foam 3), smooth foam (Foam 2), cardboard (Paper 5), and fabric (Fabric 5). a) Pressing force as a function of time used for softness rendering. b) Skin temperature over time used for thermal rendering. c) Top view images of surfaces used for roughness rendering.}
    \label{fig:ChosenTexturesPhoto}
\end{figure*}

The SENS3 database contains force sensor recordings of participants pressing materials with their index fingers while mounted on a linear stage. During each pressing action, the linear stage moved vertically at a constant speed (2~mm/s) until the applied force reached 3~N. Hence, the target force was achieved more slowly in low stiffness materials than in stiffer ones, in which it was reached quickly. Figure~\ref{fig:ChosenTexturesPhoto}a shows the measured pressing and lifting data in the SENS3 database for selected surfaces. See Section~\ref{sec:experiment} and \ref{app2} for the detailed reasoning for selecting these surfaces.

We leveraged these force profiles to render softness by calculating the pressing rate from the recorded data and mapping it to the actuator's inflation speed. The moment of peak force in each signal was identified to mark the transition from the pressing phase to static contact. From these profiles, we computed the linear pressing slope for each texture by measuring the time interval between the initial detected force and the peak force. These pressing slopes were then mapped onto the actuator’s speed range: the softest texture (rough foam) corresponded to the minimum inflation speed, while the stiffest texture (rough metal) was assigned to the maximum. This mapping used the complete actuator range, enabling dynamic inflation rates for more realistic softness rendering.

In this mapping, materials with a lower pressing slope--indicating softness--produced slower inflation speeds and lower applied forces. Conversely, stiffer materials with steeper pressing slopes inflated more quickly, generating stronger forces. A similar procedure was used to determine deflation speed by identifying the point at which the measured force began to decrease, signaling finger lift-off. Softer materials were rendered with gradual deflation, whereas stiffer materials produced faster, sharper deflation profiles. 

\begin{figure}[!t]
    \centering
    \includegraphics[width=\textwidth]{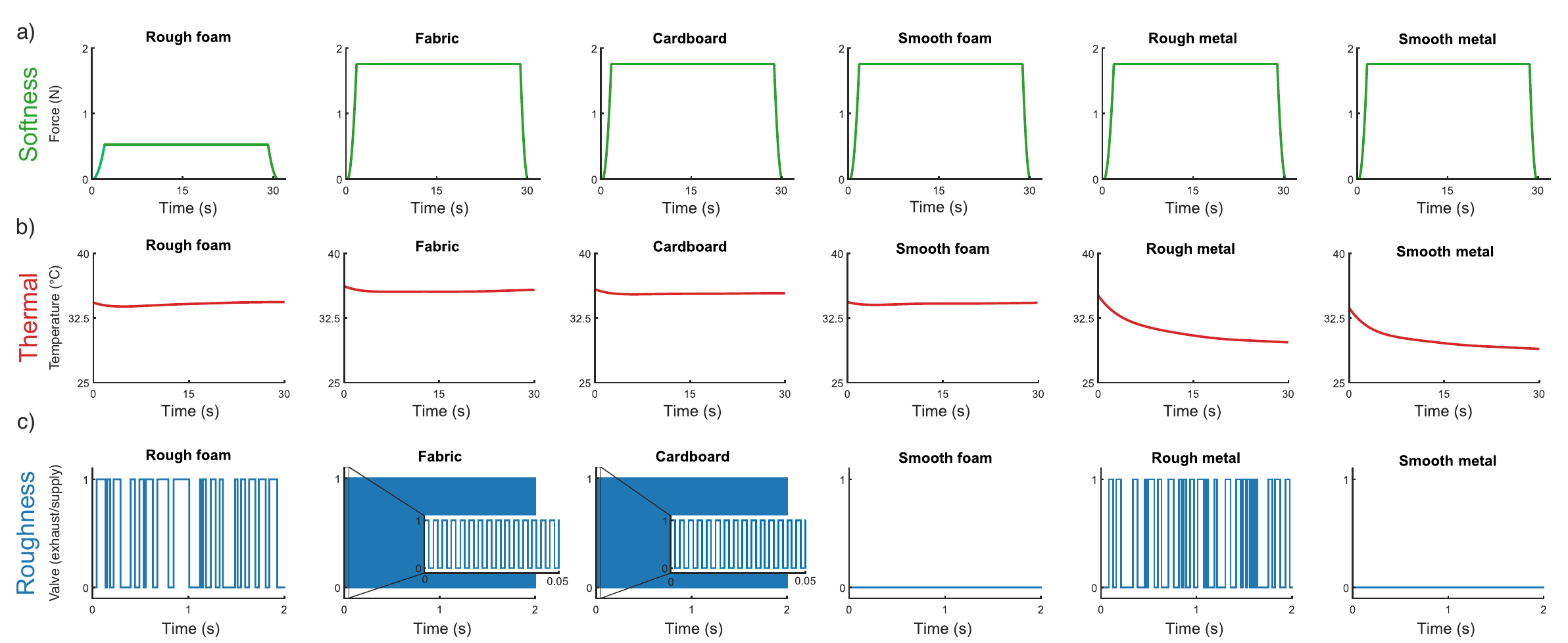}
    \caption{The signals sent to our soft haptic ring to render the selected textures. a) Displacement of the linear actuator connected to the pneumatic pouch for softness rendering. b) Display (hydraulic tube) temperatures for thermal rendering. c) Binary valve (fast switching solenoid) operation commands for roughness rendering. While the roughness rendering lasted 10 seconds during the experiments, only the first 2 seconds of the data are visualized for clarity. }
    \label{fig:rendering}
\end{figure}

As a result, the rendering signal follows a trapezoidal profile (see Figure~\ref{fig:rendering}a), where the rising and falling slopes correspond to the actuator’s inflation and deflation speeds, respectively. During the static contact phase, the rendering signal forms a plateau, maintaining a constant applied force. While the actual pressing signals in the database exhibit a slight decrease in force during this interval, we opted to simplify our model due to the limited controllability of our syringe-linear actuator setup. The actuator remains inflated and provides a steady force by keeping the linear actuator stationary during this phase.

\subsubsection*{Thermal Rendering (Static Contact Action):}
Our perception of an object's temperature, whether warm or cool, depends on the change in skin temperature caused by the heat transfer between the object and our skin. Typically, the temperature of the hand varies between 25~$^\circ$C and 36~$^\circ$C \citep{verrillo1998_hydration}. Hence, when one touches an object at room temperature (20-22~$^\circ$C), heat transfer from the skin to the object occurs. The change in the skin's temperature at the contact point depends on the initial temperatures of the skin and material as well as the thermal properties of the material \citep{klatzky2013_material}. 

Rendering of thermal response between the skin and an object is typically achieved by modeling and replicating the heat transfer between the skin and the desired object. Many models have been proposed in literature, including the works by \cite{johnson1979_thermal} and \cite{Jones2008}, which model the contact between skin and material as stationary. 

To render thermal cues that replicate the heat transfer between the skin and a surface, we used the semi-infinite body model proposed by \cite{Ho2006, Ho2008}. Simulating object-skin interaction temperatures with a thermal display using this model requires knowledge of the skin temperature upon contact, the heat flux, and the thermal contact resistance between the skin and the display. The skin temperature and heat flux data during static contact were retrieved from the SENS3 database, as seen in Figure~\ref{fig:ChosenTexturesPhoto}b. We estimated the contact resistance of the designed display, in this case the silicone tube embedded in the ring, as 0.0015~$m^2K/W$~\citep{SiliconeCoefficient}. See \ref{app:semi-infinite-body-model} for more details on implementing the semi-infinite body model.

We pre-processed the skin temperature and heat flux data by applying a 10~Hz and a 1~Hz low-pass filter, respectively, to remove high-frequency noise. We then discarded the initial portion of each recording in which the finger had not yet contacted the surface, as the model is only valid during skin-object interaction. After these steps, we computed the display temperature and reduced its complexity by fitting a seventh-order polynomial to the resulting curve (Figure~\ref{fig:rendering}b). This processed temperature profile was sent to the microcontroller, which actuated the hydraulic circuit to pump water through the silicone tube to reproduce the desired thermal transitions. Thermal rendering was synchronized with the static contact actions, but it was disabled during sliding, as humans are unable to reliably discriminate surface temperature during active exploration~\citep{green2009_active, peters2023_thermal}. For the materials rendered in this work, the largest required thermal change was approximately 4 $^\circ$C over 15 s, which can be achieved by the ring (Figure~\ref{fig:temperature}) when initialized to a baseline temperature of approximately 35 $^\circ$C.

\subsubsection*{Roughness Rendering (Slide Action):}
The perceived roughness of objects is influenced by skin deformations when pressed against surfaces, as well as vibrations and stretch when the skin moves across them~\citep{weber2013_textures}. Various methodologies and haptic devices have been proposed to render surface roughness. One common approach models interaction signals from real surfaces, such as forces or accelerations, in the frequency domain and reproduces them using vibrotactile or surface haptic devices~\citep{romano2012_texture, fiedler2019_texture}. Other methods modulate interaction forces or vibrations by computing local gradients of a grayscale image of the surface~\citep{basdogan1997_ray, kim2013_3D}. While the first approach enhances realism by incorporating contact interactions, the second effectively conveys surface topography.

Given the bandwidth constraints of our actuator (30-300~Hz), we adopted an image-based approach to render surface roughness. Our method combines an intensity-based algorithm with a peak detection function applied to high-resolution surface images. This approach leverages the relationship between surface texture and intensity variation: smoother textures exhibit more uniform intensities, while rougher textures show greater variance due to features such as bumps and creases.

To implement this method, we first applied a mean filter to grayscale surface images, reducing noise and eliminating high-frequency components that our valve cannot reproduce. Next, assuming uniform texture distribution, we extracted a one-dimensional signal by selecting a horizontal line of pixels at the texture’s mid-height. We then employed MATLAB’s findpeaks function~\citep{findpeaks} to analyze the signal’s relative intensity fluctuations. This function identified local maxima and minima based on neighboring values, enforcing a minimum separation distance aligned with the valve’s operating frequency. Peaks were selected based on their prominence (height relative to surrounding values), where local maxima indicated deactivation points for the valve, and local minima signaled activation.

The algorithm’s output was a binary square wave (see Figure~\ref{fig:rendering}c), structured according to pixel positions. To translate this into a time-based control signal, we applied a conversion factor of 50 mm/s, a speed chosen to balance dynamic exploration while operating within the valve’s limitations.

Despite the advantages of our peak detection algorithm, it has limitations when processing surfaces with very high- or low-frequency textures, such as fabrics and cardboard, or smooth metal, respectively. High-frequency textures produce intensity signals with small, densely packed peaks. When the algorithm fails to distinguish these peaks, it underestimates the spatial frequency, causing the texture to be interpreted as coarser than it is. To address this issue, we manually adjusted the square wave frequencies for these fine textures, setting them to the valve’s maximum operating frequency (300~Hz). At frequencies below 30~Hz, the valve may remain open for extended periods, increasing the risk of over-inflation and potential rupture of the pneumatic pouch. To prevent this, we ensured that each actuation signal limited the valve’s open duration to a maximum of 1.67~ms.

\section{User Study}\label{sec:experiment}
The user study aimed to assess the effectiveness of our haptic ring and texture rendering methodology in simulating the multimodal properties of real textures.

\subsection*{Participants:}
Five women and ten men with an average age of 25.33 years (standard deviation, SD: 2.18) participated in the user study. Only one participant was left-handed. None of them presented any visual or sensory-motor disabilities. Regarding ring size, only one participant used the small-sized ring, five used the medium-sized ring, and nine employed the large-sized ring. The experimental protocol was approved by the TU Delft Human Research Ethics Committee (approval number 4546). All participants gave their informed consent.

\subsection*{Stimuli: }
Our stimuli consisted of six textures---rough metal, smooth metal, rough foam, smooth foam, cardboard, and fabric---corresponding to Metal~1, Metal~5, Foam~3, Foam~2, Paper~5 and Fabric~5 from the SENS3 database (Figure~\ref{fig:ChosenTexturesPhoto}) and their virtual counterparts generated via the explained rendering methodology (Figure~\ref{fig:rendering}). These textures were chosen for their distinct thermal, roughness, and softness properties, aligning with our device’s actuation capabilities along these perceptual dimensions. The selection was guided by the reported principal component and factor loading analysis results of the adjective rating experiment conducted by~\cite{Balasubramanian2024} across the 50 textures in the database. Figure~\ref{fig:PCA} visualizes the distribution of these textures in the perceptual space after the principal component and factor loading analysis. See ~\ref{sec:texture_selection} for a more in-depth explanation of the selection procedure for these textures. 

\subsection*{Experimental Setup:}
\begin{figure}[h]
    \centering
    \includegraphics[width=0.48\textwidth]{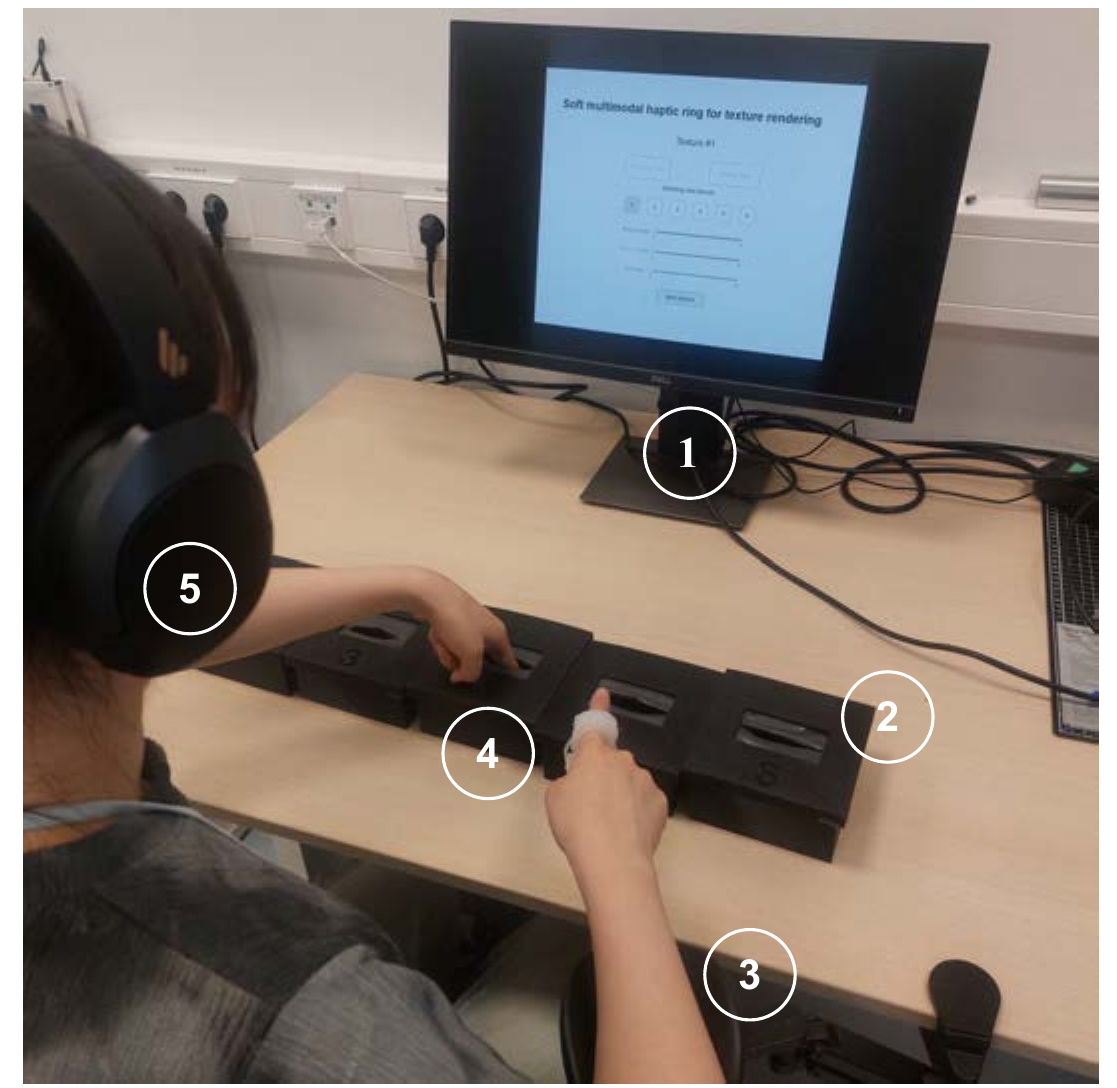}
    \caption{Experimental setup: (1) monitor displaying GUI, (2) textures within the 3D-printed enclosure, (3) armrest, (4) soft multimodal haptic ring, (5) noise-canceling headphones}
    \label{fig:ExperimentalSetup}
\end{figure}

Each participant sat in front of a monitor and a set of six textures during the experiment (Figure~\ref{fig:ExperimentalSetup}). Each texture was enclosed and hidden from the participant's view by a custom 3D-printed case. The cases had openings covered by a plastic layer obscuring the textures visually while allowing participants to explore them freely by touch. Each box was labeled with a number from one to six to facilitate texture selection. Participants provided their responses through a graphical user interface (GUI) displayed on the monitor, using a computer mouse to interact with the system.

Participants wore the soft multimodal ring on their dominant hand's proximal phalanx of the index finger. For added comfort, they could rest their dominant arm on an armrest positioned in front of the textures. Throughout the experiment, users wore noise-canceling headphones (W830NB, Edifier) that played white noise to block auditory cues and minimize distractions.

\subsection*{Experimental Procedure:}
Our experiments consisted of two tasks: a rating task and a matching task. In the rating task, participants rated the perceived sensations of presented textures (real or virtual) using the adjective pairs flat–bumpy, cold–hot, and soft–stiff. These pairs were selected as they are commonly used in literature to represent texture's roughness, thermal, and softness dimensions~\citep{Okamoto2013}. In the matching task, participants matched the virtual textures experienced through the ring with the six selected real textures. 

Before the experiments began, each participant thoroughly washed their hands. Then, they were briefed about the study and asked to explore each visually-occluded physical texture using their dominant hand's index finger. When exploring the surfaces, they were asked to apply the press, static contact, and sliding actions (not necessarily consecutively). During this process, participants were presented with a graphical user interface (GUI), where they were asked to rate each texture based on flat-bumpy, cold-hot, and soft-stiff adjective pairs using the sliders displayed on the GUI. Each slider had 100 increments for each adjective pair. They were allowed to switch between the physical textures and explore them as much as they wanted. They were instructed to rate the textures considering only the six samples presented. 

After completing the first adjective rating stage, participants wore the multimodal ring on the index finger of their dominant hand. The experimenter ensured the proper placement of the actuators and the secure positioning of the sensing elements. The second part of the experiment commenced once participants were familiar with the setup. During this stage, participants were asked to interact with a GUI displaying two actions: press-wait-lift and slide. The press-wait-lift action simulated pressing and maintaining static contact with a surface for 30 seconds, while the slide action simulated the lateral exploration of a surface at a speed of 50~mm/s for 10 seconds. The GUI displayed a virtual ball moving at the corresponding rates and directions to guide participants' exploratory action. For the sliding action, the ball moved horizontally, while for the pressing action, the ball moved vertically, descending and then remaining stationary to represent static contact. The participants were instructed to mimic these movements in mid-air with their ring-equipped index finger. They were also encouraged to continue exploring the real textures with their non-dominant hand as much as needed throughout the experiment. 

Participants were instructed to start every texture exploration with the slide action. Upon pressing the corresponding button in the GUI, a five-second countdown began, followed by the appearance of the moving ball on the screen and the start of the roughness simulation in which vibratory stimuli were delivered. After completing the sliding simulation, participants proceeded to the press-and-wait action, where temperature and pressure cues were provided simultaneously. For this action, the countdown did not start immediately after pressing the button. Instead, a "preparing" label appeared on the screen to indicate a preparation stage, during which the water and ring tube display were heated to the required initial temperature. Once this temperature was reached, the five-second countdown began, followed by the appearance of the moving ball on the screen. 
 
After completing both actions, participants were asked to select the real surface that best matched the virtual texture. They also rated the perceived virtual stimuli using the previously mentioned adjective pairs. Both tasks were completed directly through the GUI. After submitting their responses, participants continued to the next texture, for which the same approach was repeated. In total, four rounds were conducted, displaying each texture once per round and accounting for a total of twenty-four trials per participant. The order of the displayed textures was randomized for each round. The first round was considered the training phase in which participants familiarized themselves with the setup. Hence, the participant responses from this phase were excluded from analysis. Participants were never informed whether they had correctly matched the virtual and physical textures.

\subsection{Results} \label{sec:results}
\begin{figure}[h]
    \centering
    \includegraphics[width=\textwidth ]{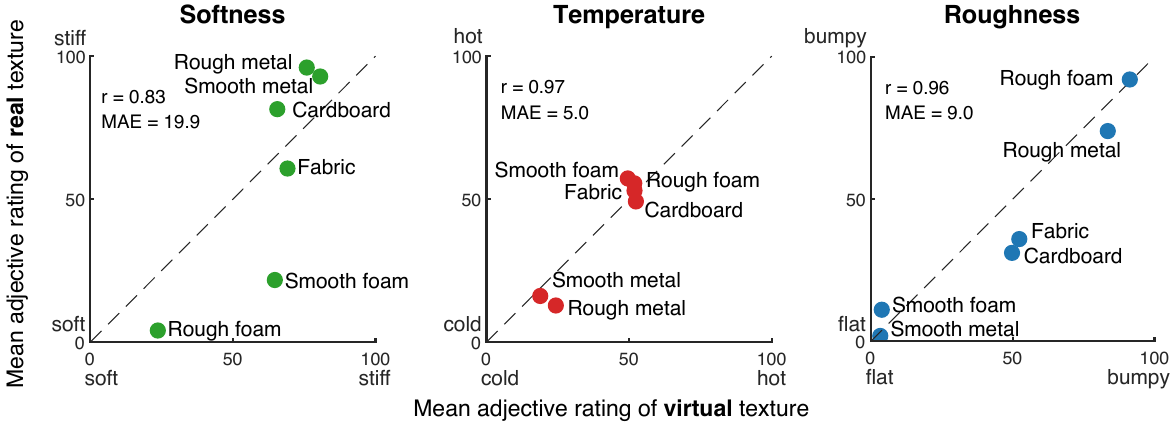}
    \caption{Comparison of mean virtual and real adjective ratings across six textures and three perceptual dimensions with Pearson coefficient (r) and the Mean Absolute Error (MAE).}\label{fig:virtualadjectiverating_correlation_summary}
\end{figure}

Figure~\ref{fig:virtualadjectiverating_correlation_summary} presents the mean adjective ratings of virtual textures compared to their real counterparts in terms of perceived roughness, temperature, and softness. Since the rating tasks for real and virtual textures were conducted in separate experiments, this summary plot offers insights into the device's rendering performance across different textures rather than serving as a direct one-to-one comparison. It provides a representation of agreement at the texture level to asses how well each modality is reproduced on average. For each modality, we report the Pearson correlation coefficient (r) to asses whether relative ranking is preserved and the Mean Absolute Error (MAE) to quantify absolute accuracy.
Detailed box plots of all virtual-real adjective rating comparisons can be found in Figure~\ref{fig:AdjectiveRatingComparison}.

\begin{figure*}[h]
    \centering
    \includegraphics[width=\textwidth]{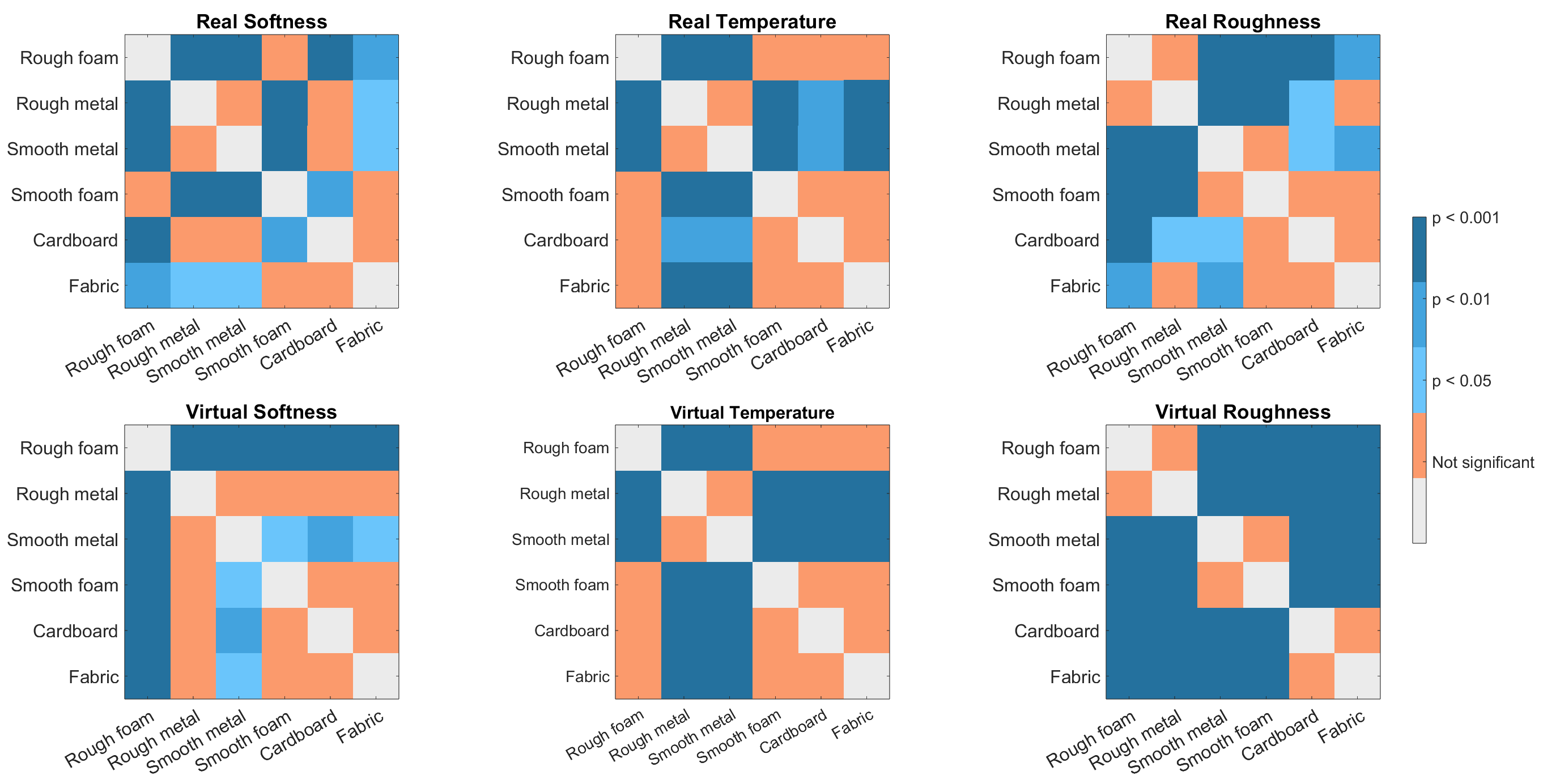}
    \caption{Results of the Kruskal-Wallis test on perceived differences in the roughness, temperature, and softness dimensions. The upper row presents the results for the real textures, while the lower row shows the results for the displayed virtual textures.}
    \label{fig:virtualadjectiverating_statistics}
\end{figure*}

Additionally, we assessed the device's rendering performance for generating variety of texture sensations. A Kolmogorov-Smirnov test was first conducted to determine the normality of each distribution, revealing significant deviations from normality (p $<$ 0.05). As a result, a Kruskal-Wallis test was performed to identify statistically significant differences in adjective ratings across textures and modalities; see Figure~\ref{fig:virtualadjectiverating_statistics}.

\begin{figure}[h]
    \centering
    \includegraphics[width=0.8\linewidth]{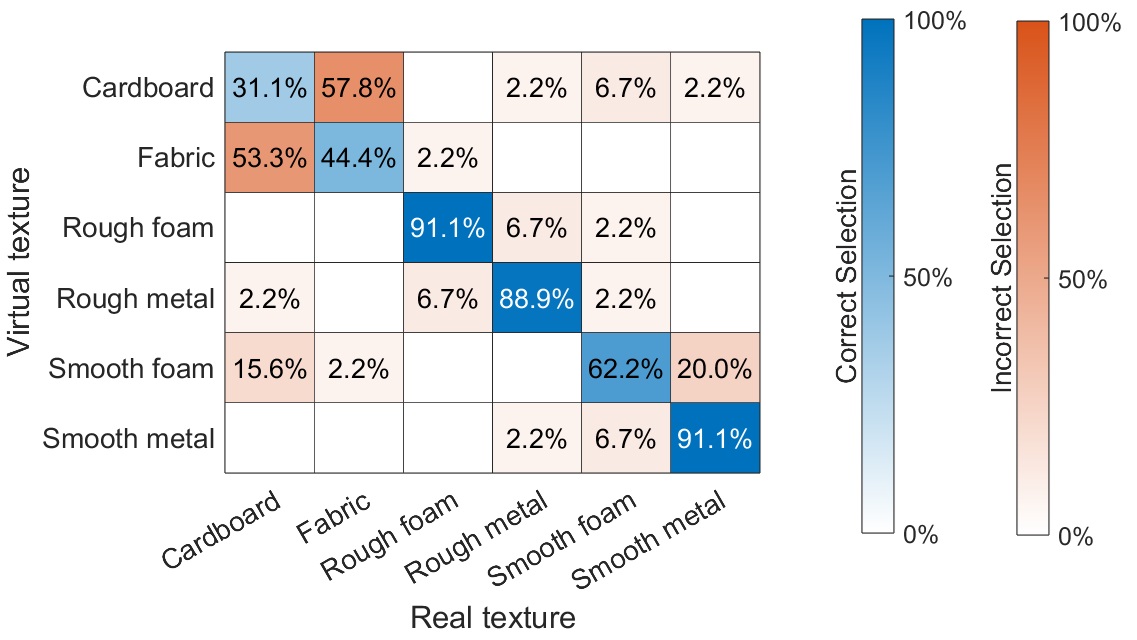}
    \caption{Confusion matrix for the texture matching task. Rows present virtual textures and columns represent selected real textures. Each cell indicates the proportion of times a given pair was chosen, with the diagonal representing the correct matches. Overall precision = 0.68, Recall = 0.69, and $F_1$-Score = 0.68.}
    \label{fig:ConfusionMatrix}
\end{figure}

The matching percentage between the virtual textures displayed by the ring and the real surfaces selected by the participants is shown in Figure~\ref{fig:ConfusionMatrix}. In this analysis, each participant's responses across the three rounds were treated as independent data points. A two-sample Pearson Chi-Squared test ($\chi^2$ (25, N = 270) = 756, p $<$ 0.001) indicated that the distribution of participant predictions significantly deviates from the expected chance level (1/6 probability per texture). The overall precision, recall and $F_1$-score are 0.6815, 0.6853, and 0.6834, respectively.

\clearpage
\section{Discussion}\label{sec:discussion}

In this paper, we introduced a soft, wearable haptic ring that delivers three fundamental tactile cues—vibration, pressure, and temperature—to reproduce texture sensations on the user’s proximal phalanx. Through systematic characterization, we demonstrated that the device reliably generates stimuli associated with perceptual roughness (smoothness), softness (hardness), and warmth (coldness). To the best of our knowledge, this is the first soft, multimodal haptic ring to integrate these three modalities within a compact, wearable form factor. 

Beyond the haptic device, we proposed a data-driven, action-based rendering approach that maps natural exploratory behaviors to the most perceptually informative texture dimensions. Rather than attempting to replicate full tactile signals, our method selectively triggers feedback based on how humans probe surfaces, prioritizing perceptual relevance over physical fidelity. Importantly, this rendering strategy considers the multimodality of texture perception for the first time in a wearable ring interface and can be adapted to other haptic devices. We validated both the device and the rendering method through a user study in which participants matched virtual textures to real materials and rated the resulting sensations.

Collectively, these contributions represent an initial step toward understanding the feasibility of transferring multimodal haptic cues from the fingertip to other regions of the hand and provide both a hardware and software foundation for achieving this goal. By leaving the fingertip unobstructed, the ring could potentially allow users to physically interact with their surroundings while perceiving superimposed virtual textures, opening new opportunities for more immersive interaction in extended reality applications. In the following paragraphs, we discuss our findings in greater detail and examine their implications.

\subsection*{Texture Reproduction Capabilities of the Soft, Multimodal Haptic Ring and Action-Based Rendering}

The characterization experiments demonstrated that the ring is capable of reliably delivering multimodal tactile cues---pressure, vibration, and temperature---within ranges relevant to human texture perception. The perceptual study further demonstrated that these cues can be meaningfully combined through our action-based rendering approach to evoke distinguishable surface sensations, indicating that the device and rendering method jointly support the reproduction of multimodal textures.

The device can generate normal forces up to 1.75~N on the skin, with a monotonic and well-controlled increase as the linear actuator advances (Figure~\ref{fig:forces}). Within this operating range, the pneumatic actuator produces clearly perceivable normal sensations, exceeding reported detection thresholds for the proximal phalanx (0.046~mN measured using a pressure aesthesiometer by \cite{weinstein1962_phalange}). Despite this capability, softness ratings of virtual textures only moderately matched judgments for their real counterparts (r = 0.83, MAE = 19.9; Figure~\ref{fig:virtualadjectiverating_correlation_summary}). Only one of the virtual foam samples was perceived as soft, whereas both foam materials were judged similarly soft in the real texture set. The real metals were consistently judged stiffer than their virtual counterparts. 

These differences could be caused by three key limitations of our approach. First, all simulated textures were reproduced through a compliant silicone interface that cannot fully convey high stiffness. Second, the ring can only provide cutaneous cues, whereas participants interacting with real surfaces received both cutaneous and kinesthetic cues---known to contribute strongly to perceived stiffness and compliance~\citep{Srinivasan1995TactualDO, Dhong2019}. Finally, softness was encoded by mapping the rate of force–displacement curves at the fingertip to actuator pressure delivered to the proximal phalanx, a transformation that can attenuate differences between materials. Together, these factors could have reduced the perceptual variability in rendered softness, with only the virtual rough foam and smooth metal rated as significantly distinct from the other textures (Figure~\ref{fig:virtualadjectiverating_statistics}).

The same pneumatic actuator can generate displacements between 0.27~mm and 0.003~mm over a 30–300~Hz frequency range (Figure \ref{fig:magnitude_response}), falling well within the perceivable range of human vibrotactile sensitivity. These values remain more than an order of magnitude larger than psychophysical thresholds reported at the fingertip for similar frequencies \citep{morioka2008_threshold, verrillo197_threshold}, suggesting that vibrotactile cues are unlikely to be limited by actuator intensity. However, unlike voice-coil or piezoelectric actuators, the pneumatic mechanism cannot generate displacements outside of this frequency range, nor can it reproduce complex vibrotactile signals in which amplitude varies over time or is independently controlled at a fixed frequency. Hence, the amplitude of the roughness variations in the virtual textures did not fully represent their real counterparts.

Despite this restricted actuation space, roughness ratings for the virtual textures closely matched those of their real counterparts (r = 0.96, MAE = 9.0; Figure~\ref{fig:virtualadjectiverating_correlation_summary}). In both virtual and real conditions, three distinct roughness levels emerged: rough foam and rough metal were perceived as highly bumpy, smooth foam and smooth metal as flat, and fabric and cardboard as moderately rough. Notably, differences were less pronounced in the virtual textures than with the real materials (Figure~\ref{fig:virtualadjectiverating_statistics}). This effect may stem from the actuator’s limited ability to render fine variations in amplitude and friction—two primary contributors to perceived roughness \citep{isleyen2020_roughness, friesen2021_space, learn2feel}—resulting in reduced perceptual contrast between virtual textures.

The thermal actuator can change the display temperature by up to 25$^\circ$C within 65~s (Figure~\ref{fig:temperature}). Although humans can detect very small temperature differences as small as 0.02$^\circ$C for cooling and 0.03$^\circ$C for warming, these thresholds are strongly dependent on the rate of temperature change \citep{Jones2008, stevens1998_thermalthreshold}. The system's relatively slow response therefore constraints the reproduction of small, rapidly varying thermal cues. Even so, the actuator’s performance was sufficient to render clearly distinguishable thermal differences between surfaces, with virtual ratings closely matching those of real surfaces (r = 0.97, MAE = 5.0; Figure~\ref{fig:virtualadjectiverating_correlation_summary}). 

This high accuracy was achieved by implementing the algorithm proposed by \citet{Ho2008}, which compensates for thermal conduction losses so that the heat flux delivered to the skin more closely matches the intended cue. Moreover, the high sensitivity to cold and
warm stimuli on the dorsal side of the hand, where the thermal stimuli were applied, may have contributed to these
results \citep{wakolbinger2014}. As in the real surfaces, two perceptual clusters emerged: metals formed a colder group, while all other materials were perceived as warmer and near room temperature. The rendered textures also exhibited comparable variability within each cluster relative to the real materials (Figure~\ref{fig:virtualadjectiverating_statistics}).  

The results of the matching task indicate that participants were able to associate the virtual textures displayed by the haptic ring with their real counterparts, as the correct matching rate for all textures was significantly higher than chance (Figure~\ref{fig:ConfusionMatrix}). The rough foam, rough metal, and smooth metal textures achieved the highest precision, reaching 90\%. Interestingly, these textures also exhibited the adjective ratings at utmost margins among the selected surfaces, likely due to their distinct characteristics, such as the pronounced temperature contrast of metals, the bumpy texture of rough foam and rough metal, or the softness of rough foam. These are visible both in our experiments (Figure~\ref{fig:virtualadjectiverating_correlation_summary}) and in the perceptual space found by \cite{Balasubramanian2024}. 

Smooth foam, with a correct identification rate of 62.2\%, was often misclassified as smooth metal or cardboard. This lower precision may stem from its perceived roughness, which is similar to smooth metal, or its temperature, which resembles that of cardboard (Figures~\ref{fig:AdjectiveRatingComparison} and \ref{fig:virtualadjectiverating_statistics}). Furthermore, the virtual smooth foam's perceived softness was less distinctive than its real counterpart, potentially contributing to the misclassification. Fabric and cardboard textures yielded the poorest matching results, with correct identification rates below 50\%. While participants could recognize that the displayed texture belonged to one of these two categories, distinguishing between them was largely a matter of chance. This outcome was expected, given their highly similar profiles across the roughness and temperature dimensions in the texture-recorded signals (Figure~\ref{fig:ChosenTexturesPhoto}). This similarity is also reflected in the adjective ratings, with only the softness dimension showing modest differences between the two textures (Figure~\ref{fig:virtualadjectiverating_statistics}). When these softness differences are translated into inflation and deflation rates, they become subtle (Figure~\ref{fig:rendering}a), falling below the ring's softness rendering resolution. 

Taken together, these results show that while each modality has distinct limitations, the combination of pressure, vibration, and temperature delivered through the action-based rendering approach enables a set of virtual textures that are both discriminable and broadly aligned with real-surface percepts. Softness rendering was limited by the compliant interface and lack of kinesthetic feedback, vibrotactile cues reproduced surface roughness with high fidelity despite a constrained signal space, and thermal cues were rendered with high perceptual accuracy even with slow actuation dynamics. These findings suggest that realistic texture perception does not require high-fidelity replication within any single modality; rather, it emerges from the complementary integration of imperfect but informative cues. This synergy points to a promising direction for wearable haptic devices, where combining multimodal cues may be more effective than maximizing fidelity within individual channels.

Additionally, the post-experiment interviews reinforced the importance of multimodality. Participants emphasized that experiencing all three haptic modalities was essential for accurately matching the perceived cues to real textures, noting that omitting any modality would have greatly increased task difficulty. The results also highlighted the prominent role of thermal cues: many participants relied more on temperature information than on softness when identifying textures, and those who experienced weaker thermal sensations reported the simulation as less realistic. This effect was particularly noticeable in one participant with a small scar near the temperature-sensing area, who described the task as ``pattern matching''  rather than  a texture experience.

\subsection*{Transfer of Fingertip Cues to the Proximal Phalanx}
This study represents an initial step toward assessing both the feasibility and constraints of transferring multimodal haptic cues from the fingertip to other regions of the hand. Our findings indicate that haptic information can be relocated to the phalanx with reasonable accuracy, as participants were able to match textures with notable precision. However, this conclusion should be interpreted cautiously given the narrow scope of the investigation.

The matching task was conducted with a limited texture set, which may have contributed to both the high matching accuracy and the similarity observed in the adjective ratings. Only six textures were included, selected to span distinct levels of roughness, softness, and temperature while minimizing participant fatigue. Although this strategy reduced experimental burden, it also restricted the diversity of perceptual characteristics being evaluated. As a result, the experiment may have captured only a subset of the perceptual challenges associated with relocating haptic cues. A broader and more continuous sampling of textures would provide a more comprehensive assessment of the ring’s rendering capabilities and could reveal breakdowns or ambiguities that remain undetectable within the current limited dataset.

In addition, although texture matching and adjective ratings provide useful evidence that haptic cues can be perceived away from the fingertip, they do not fully capture how much information is preserved through relocation. Future studies should assess how accurately relocated cues convey spatial and temporal detail, intensity, and overall realism to determine the fidelity of haptic transfer to other regions of the hand.

Another limitation of the current design is that stimulation is restricted to a single site per modality, which may constrain the fidelity of the rendered cues. For thermal feedback, a single heat source reduces the surface’s thermal representation to a single value; however, this constraint is likely to have limited perceptual impact, as humans exhibit low thermal spatial resolution and tend to integrate temperature over an area \citep{kappersThermalPerceptionThermal2019, hoMaterialRecognitionBased2018}. In contrast, roughness rendering may be more affected by this restriction, as reproducing textures or surface shapes that rely on spatial variations in pressure or vibration is inherently limited with single-point actuation. Prior work demonstrates that spatial variation in skin indentation remains perceptually relevant even for fine textures~\citep{weber2013_textures, grigoriiSpatialProfileSkin2022b}. This suggests that future ring designs could benefit from multi-site or ``multi-pixel'' haptic rendering, such as arrays of small inflatable pouches~\citep{tan2025_wearable}. However, it remains unclear to what extent increased spatial resolution would enhance perception at relocated sites with lower spatial acuity, and how combining multiple modalities within a confined area might affect cue integration. Characterizing these interactions may be critical for achieving high-fidelity haptic rendering in compact wearable systems.

Finally, despite these limitations, our observations align with prior work showing that humans can discriminate texture information through ring-based displays \citep{Gaudeni2019, Friesen2024, normand2025_augment}. Together, these results support the potential for relocating haptic feedback away from the fingertip, while underscoring the need for more in-depth evaluation of cue salience, spatial resolution, and control fidelity before this approach can be considered a robust alternative to fingertip stimulation.

\subsection*{Future Recommendations}

Despite the careful manufacturing process, our device had some limitations. For instance, while we produced rings in three sizes to accommodate various finger types, the fit was not always perfect. Some participants found them slightly too tight or loose, potentially affecting the perception of displayed cues~\citep{pra2023pressure, galie2009contact}. Additionally, despite efforts to maintain consistent inflation characteristics across all rings, minor variations in the manufacturing process may have introduced slight differences in performance. Nevertheless, when comparing a newly produced ring with a year-old ring, no noticeable shift in their mechanical responses and dynamic behavior was observed, indicating that there was no noticeable degradation in the silicone with time. Moving forward, standardizing fabrication procedures and incorporating an adjustable design---potentially informed by individualized squeeze-force measurements as in \citet{Friesen2024}---may improve both consistency and user fit.

Although the proposed ring is soft, lightweight, flexible, and wearable, its operation currently relies on bulky external hardware, limiting its practicality in real-world settings. Despite the promising results demonstrated in this work, achieving greater portability will require simplifying the overall system, even if this comes at the expense of some softness and compliance.
A major opportunity for improvement lies in the thermal actuation mechanism. Replacing the current hydraulic system with a Peltier-based thermal interface, similar to \citep{riessenRelocatingThermalStimuli2024}, could substantially reduce system complexity and eliminate thermal delays caused by fluid transport. Although a Peltier element would introduce rigidity, this limitation could be mitigated by embedding the component within a silicone layer or a soft thermally conductive interface. Alternatively, emerging flexible thermal actuators—such as the bidirectional heating–cooling interface demonstrated by \cite{LeeJinwoo2020}—offer a promising solution that maintains softness without sacrificing control performance.
The pneumatic subsystem also presents opportunities for portability. This could be enhanced by replacing the benchtop air compressor with miniature pressure sources, such as cartridge-based systems demonstrated by \cite{Talhan2020}. Furthermore, replacing the pneumatic actuator with a piezoelectric or voice-coil actuator could not only improve the realism of roughness rendering but also enable independent control of softness and texture cues.
Finally, the current prototype does not include embedded sensing for hand motion tracking, and interaction is limited to a predefined set of sequential gestures. Future iterations could incorporate onboard sensing to enable continuous interaction.
Together, these advancements would enable a more compact, fully wearable version of the device with greater control fidelity, making multimodal texture rendering more practical for everyday interaction and extended-reality applications.

Moreover, practical deployment of relocated haptic feedback introduces perceptual challenges. In complex or visually rich interactions, the mismatch between the visible point of contact and the displaced tactile sensation can lead to spatial and temporal confusion~\citep{palmerHapticFeedbackRelocation2022a}. Even so, users generally prefer relocated feedback over the absence of tactile cues~\citep{terentiDistalHapticTouchscreensUnderstanding2025}, suggesting that this confusion may be tolerable or reduced through proper system design. Improving the synchronization of visual and haptic cues—rather than relying solely on device fidelity—may help create a more coherent multisensory percept and facilitate practical deployment of relocated haptics.

\subsection*{Potential Applications}

Our approach enables haptic textures to be relocated from the fingertip to other regions of the hand, offering new opportunities for interaction beyond traditional touch interfaces. Such augmentation can leverage the natural tactile feedback of everyday objects while adding additional sensory information, expand sensory capabilities for accessibility, support precision in dexterous or guided tasks, and enrich aesthetic or entertainment experiences, as summarized by \citet{bhatia2024_augment}. These functions can be applied broadly across domains such as communication, e-commerce, education, healthcare, art appreciation, and robotics.

\section{Data availability}
The data and codes used in this paper will be publicly available upon acceptance. 
\section{Acknowledgements}
This work was partly funded by the Dutch Research Council (NWO) with projects 19153 and OCENW.XS23.1.178 and European Research Council with project 101220242. The authors also thank Jagan K. Balasubramanian for helping with the use of the SENS3 database, Haewon Jeong for fabricating soft materials, and Khoa Lahn for his help in thermal rendering.

\section{CRediT authorship contribution statement}
\noindent\textbf{Ana Sanz Cozcolluela:} Conceptualization, Methodology, Software, Hardware, Formal Analysis, Investigation, Data Curation, Writing - Original Draft, Writing - Review \& Editing, and Visualization. \textbf{Koen W\"osten:} Methodology, Software, Hardware, Formal Analysis, Investigation, Data Curation, Writing - Review \& Editing, and Visualization. \textbf{Yasemin Vardar:} Conceptualization, Methodology, Formal Analysis, Writing - Original Draft,  Writing - Review \& Editing, Visualization, Supervision, and Funding Acquisition.
\section{Declaration of competing interest}
The authors declare they have no known competing financial interests or personal relationships that could have influenced this work.
\section{Declaration of generative AI and AI-assisted technologies in the writing process}

The authors used OpenAI's ChatGPT and Grammarly Inc.'s Grammarly tool to enhance the language and readability of this work. After using these tools, the authors carefully reviewed and edited the content as needed. They take full responsibility for the final publication.
\appendix

\section{Experimental setups used for mechanical characterization of the multimodal ring.}

\begin{figure}[h!]
    \centering
    \includegraphics[width=0.8\linewidth]{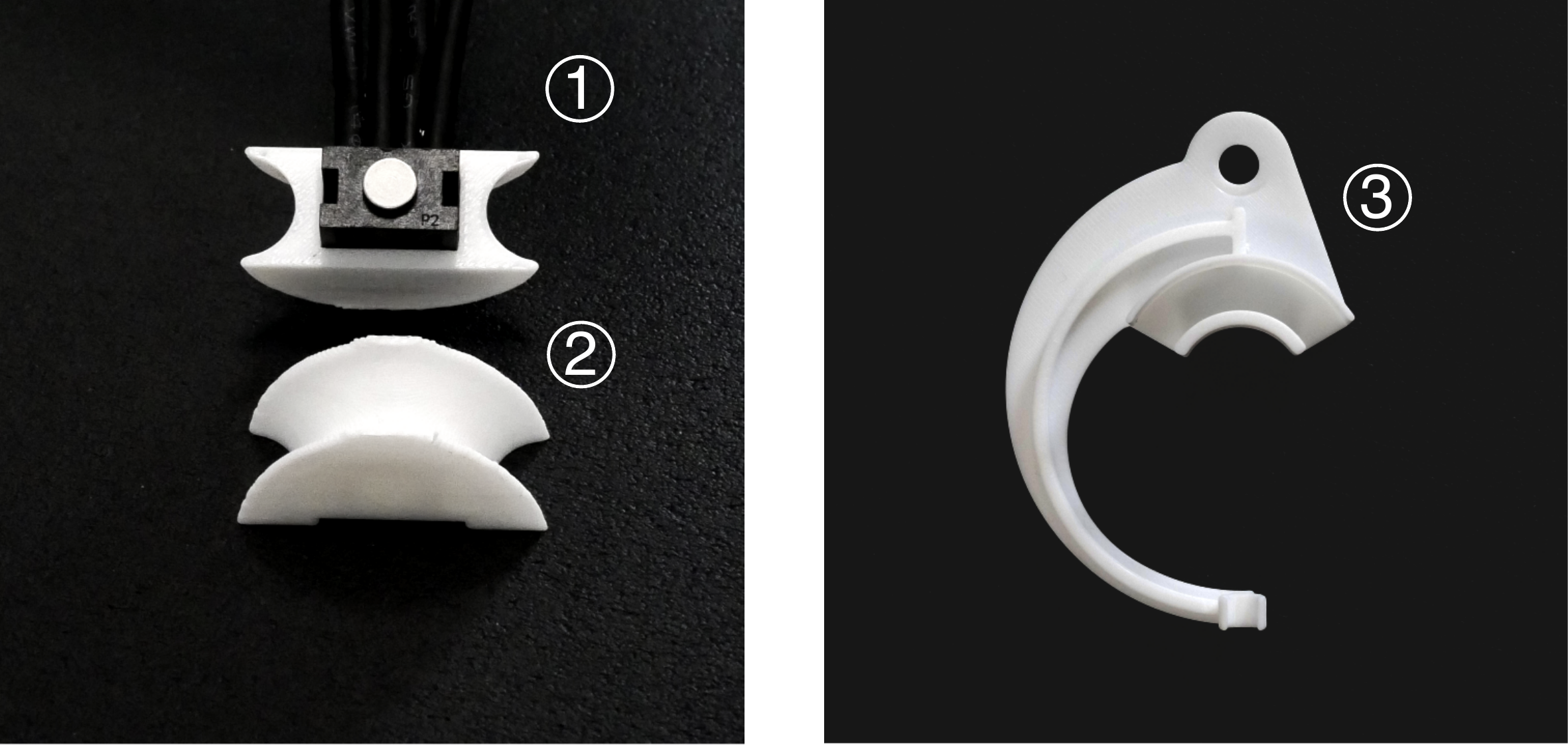}
    \caption{Force measurement unit for pressure characterization: (1) force sensor, (2) custom saddle jig to mount the sensor inside the ring, and (3) the ring holder used during displacement measurements.}
    \label{fig:force_setup}
\end{figure}

\begin{figure}[h!]
    \centering
    \includegraphics[width=0.8\linewidth]{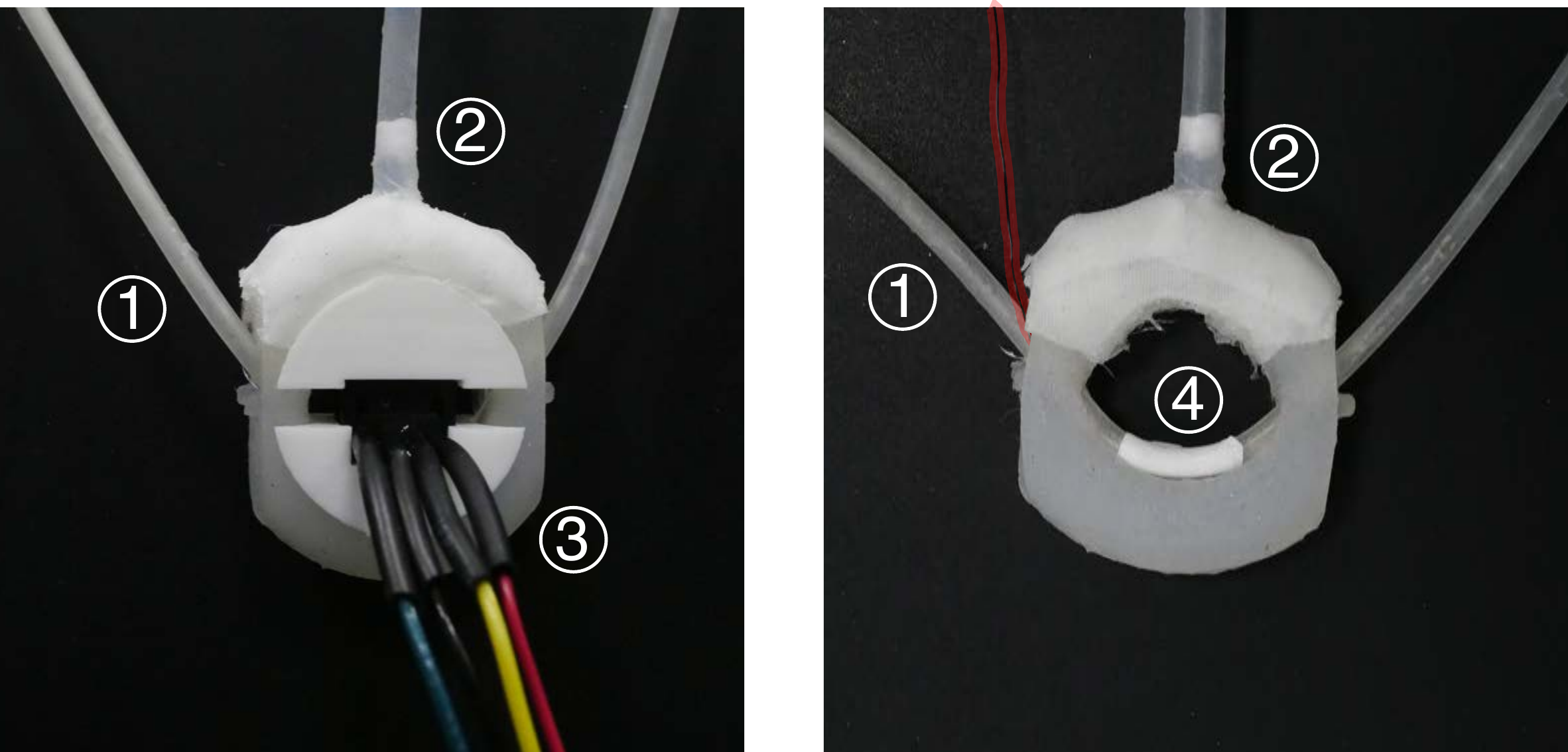}
    \caption{Measurement setups for pressure and thermal characterization: (1) silicone water tube, (2) pneumatics tube, and (3) custom saddle jig and force sensor for the pressure characterization. (4) Thermistor fixed on the silicone water tube using a winding of PTFE tape for thermal characterization.}
    \label{fig:force_thermal_setup}
\end{figure}

\begin{figure}[h!]
    \centering
    \includegraphics[width=\linewidth]{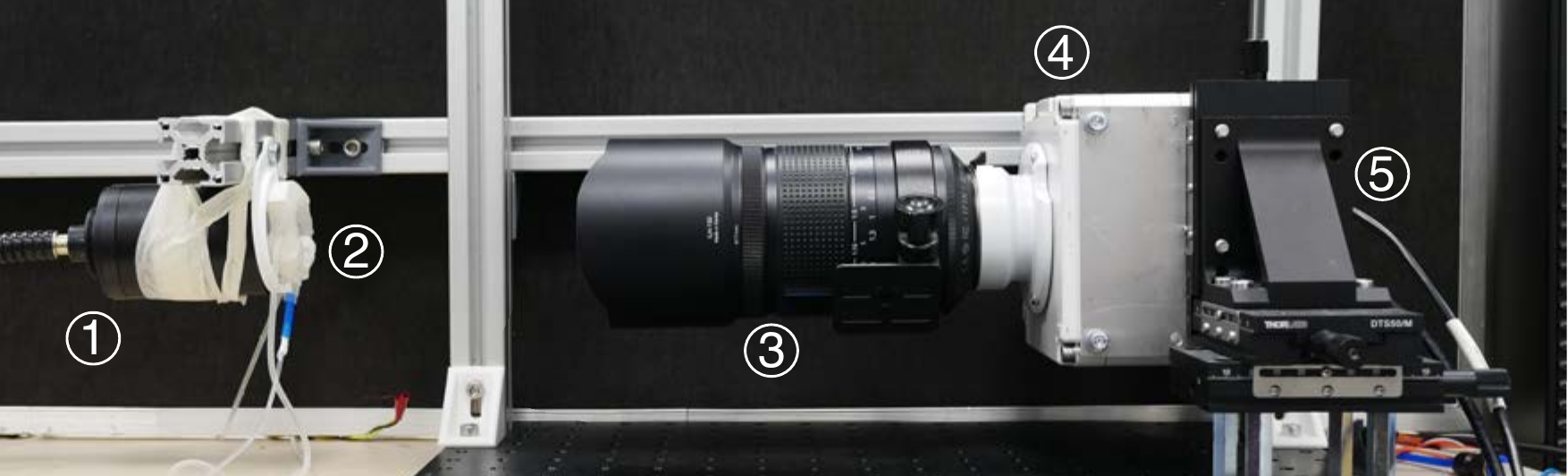}
    \caption{Experimental setup for vibration characterization: (1) light source, (2) multimodal ring under test, (3) lens, (4) camera, and (5) positioning system composed of three translation stages (DTS50/M, Thorlabs).}
    \label{fig:displacement_setup}
\end{figure}
\label{app1new}
\newpage

\section{Sub-pixel edge detection from high-speed imagery}
Displacement amplitudes were extracted from each high-speed frame by processing a one-dimensional scanline intersecting the ring edge. The initial edge position $x_0 \in \mathbb{N}$ was identified as the first pixel to drop below a manually-tuned threshold set just above the background level. For each scanline, a gradient $g[x]$ was computed using a 4th-order centered finite difference approximation.

\begin{equation}
    g[x] = \frac{-f[x+2] + 8f[x+1] - 8f[x-1] + f[x-2]}{12},
\end{equation}

where $f[x]$ denotes the pixel intensity along the scanline at position $x$.

A background noise template $g_b[x]$ on the gradient was obtained by taking the median of all scanline gradients. Global illumination trends were removed by projecting the raw gradient $g[x]$ onto the background noise $g_b[x]$ with a Gaussian weighted window $w[x]$ around $x_0$. 

\begin{equation}\label{eq:gaussian_window}
w[x] = \exp\left[-\frac{1}{2}\left(\frac{x - x_0}{\sigma}\right)^2\right], \text{ zeroed within } [x_0 - 5, x_0 + 5]
\end{equation}

\begin{equation}
\tilde{g}[x] = g[x] - \alpha g_b[x],
\qquad
\alpha = \frac{\sum_x w[x]\, g[x]\, g_b[x]}{\sum_x w[x]\, g_b^2[x]}.
\end{equation}

This yields a corrected gradient $\tilde g[x]$, and the coarse edge $x_{c}$ was then estimated as the location of the maximum value of $\tilde g[x]$ in a search window around $x_0$. 

\begin{equation}
x_{c} = \operatorname*{arg\,max}_{x \in [x_0 - \Delta,  x_0 + \Delta]} \tilde{g}[x], 
\quad \Delta \in \mathbb{N}.
\end{equation}

The final amplitudes are then estimated with sub-pixel accuracy by fitting a parabola to the three gradient samples surrounding $x_{c}$. The fractional offset $\delta$ from the central pixel was computed as
\begin{equation}
\delta = \frac{\tilde{g}[x_{c}-1] - \tilde{g}[x_{c}+1]}{2\big(\tilde{g}[x_{c}-1] - 2\tilde{g}[x_{c}] + \tilde{g}[x_{c}+1]\big)},
\quad |\delta| \le 0.5,
\end{equation}

and the final edge coordinate was
\begin{equation}
x_\text{edge} = x_{c} + \delta.
\end{equation}\label{app3}
\section{Implementation of semi-infinite body model}

The semi-infinite body model proposed by \cite{Ho2006, Ho2008} considers that the thermal interaction between the skin and the material with which it is in contact is a transient process dominated by heat conduction. In this process, heat is transferred across the interface and flows through a thermal contact resistance. Consequently, during contact, the heat flux of the skin and that of the material will be identical, where the flux is determined by the difference between the skin surface temperature and the object surface temperature divided by the thermal contact resistance. This relationship is described by Equations \ref{eq:heatflux} and \ref{eq:heatfluxformula}, where $q''_{skin,s}$, $q''_{object,s}$, $T_{skin,s}$ and $T_{skin,s}$ refer to the heat flux and temperature of the skin and object surfaces, respectively, and $R_{skin, object}$ stands for the thermal contact resistance between skin and object.

\begin{equation} \label{eq:heatflux}
q''_{skin,s} = q''_{object,s} = q''
\end{equation}

\begin{equation} \label{eq:heatfluxformula}
q''=\frac{T_{skin,s} (t) - T_{object,s}(t)} {R_{skin, object}}
\end{equation}

Thermal contact resistance depends on several variables related to the thermal and mechanical properties of the contacting surfaces. For materials with an even surface and interaction contact forces around 2~N, it can be approximated by Equation \ref{eq:thermalcontactreistance}, where $k_{object}$ is the thermal conductivity of the object.

\begin{equation} \label{eq:thermalcontactreistance}
R_{skin, object}=\frac{0.37 + k_{object}}{1870 \cdot k_{object}}
\quad \text{[$m^2K/W$]}
\end{equation}

The semi-infinite body model can be further extended to predict the temperature at which a thermal display should be activated to reproduce the thermal characteristics of an object in contact with the skin \citep{Jones2008}. This can be done simply by replacing the object terms of the previous equations with those of the display:

\begin{equation} \label{eq:heatfluxdisplay}
q''=\frac{T_{skin,s} (t) - T_{object,s}(t)}{R_{skin, object}}=\frac{T_{skin,s}(t) - T_{display}(t)}{R_{skin- display}},
\end{equation}
\begin{equation} \label{eq:displayeq1}
\begin{split}
T_{display}(t)=T_{skin,s}(t) \cdot \left[1-\frac{R_{skin,display}}{R_{skin,object}}\right]+ \\ \frac{R_{skin,display}}{R_{skin,object}}   \cdot T_{object,s}(t).
\end{split}
\end{equation}

These equations, when combined with Equation \ref{eq:heatfluxformula}, yield an expression that allows us to compute $T_{display}$ as a function of the heat flux and the thermal contact resistance between the skin and thermal display:

\begin{equation} \label{eq:displayeq2}
T_{display}(t)=T_{skin,s}(t) - q'' \cdot R_{skin-display} . 
\end{equation}\label{app:semi-infinite-body-model}

\section{Texture selection}\label{sec:texture_selection}
\begin{figure*}[h]
        \centering
    \includegraphics[width=0.8\textwidth]{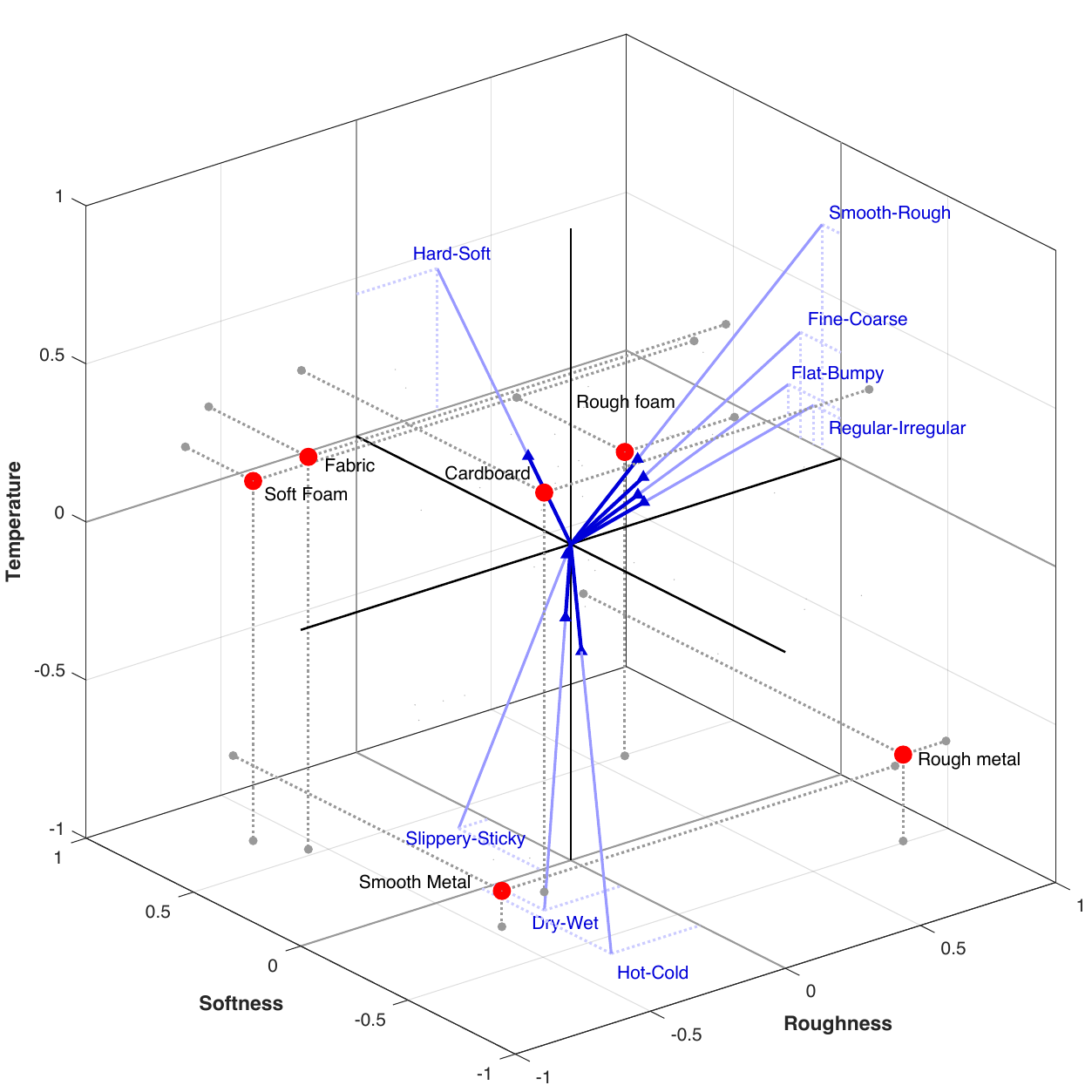}
    \caption{The locations of the selected surfaces in the perceptual space which was obtained by applying principal component and factor loading analyses on the perceptual rating experiments conducted by \cite{Balasubramanian2024}. Chosen textures are displayed as large red dots, with corresponding 2D projections shown on the cube faces.  }
    \label{fig:PCA}
\end{figure*}

We selected six textures from the SENS3 database based on the analysis conducted by \citep{Balasubramanian2024}. The selected textures were rough metal (Metal1), smooth metal (Metal5), rough foam (Foam3), smooth foam (Foam2), cardboard (Paper5), and fabric (Fabric5). In their experiments, participants interacted with 50 surfaces and rated their psychophysical sensations. Then, they used principal component analysis (PCA) to analyze the adjective ratings. From the analysis, they chose four dimensions that represented 95\% of the total variance. To enhance interpretability and ensure coverage of perceptual diversity, they performed factor analysis through varimax rotation of the component matrix. 

To choose a final set of textures, each texture was represented by its rotated component scores on the three dimensions corresponding to roughness, softness, and temperature as revealed by their rotated loadings. Figure~\ref{fig:PCA} visualizes the locations of the selected surfaces in the perceptual space.
In this three-dimensional space, textures were classified into eight octants split along the axes. Within each octant, stimuli were ranked by their distance from the origin, and the texture with the largest distance to the origin was selected as representative for that octant. From these representatives, only the five with largest distances were retained. This resulted in a set of textures diversified along interpretable perceptual dimensions aligned with the modalities renderable by the ring. Additionally, fabric (Fabric5) was included because it closely resembles soft foam (Foam2), allowing comparison between textures with subtle material differences.

\label{app2}
\newpage
\section{Comparison of virtual and real adjective ratings}

\begin{figure*}[h!]
    \centering
    \includegraphics[width=\textwidth]{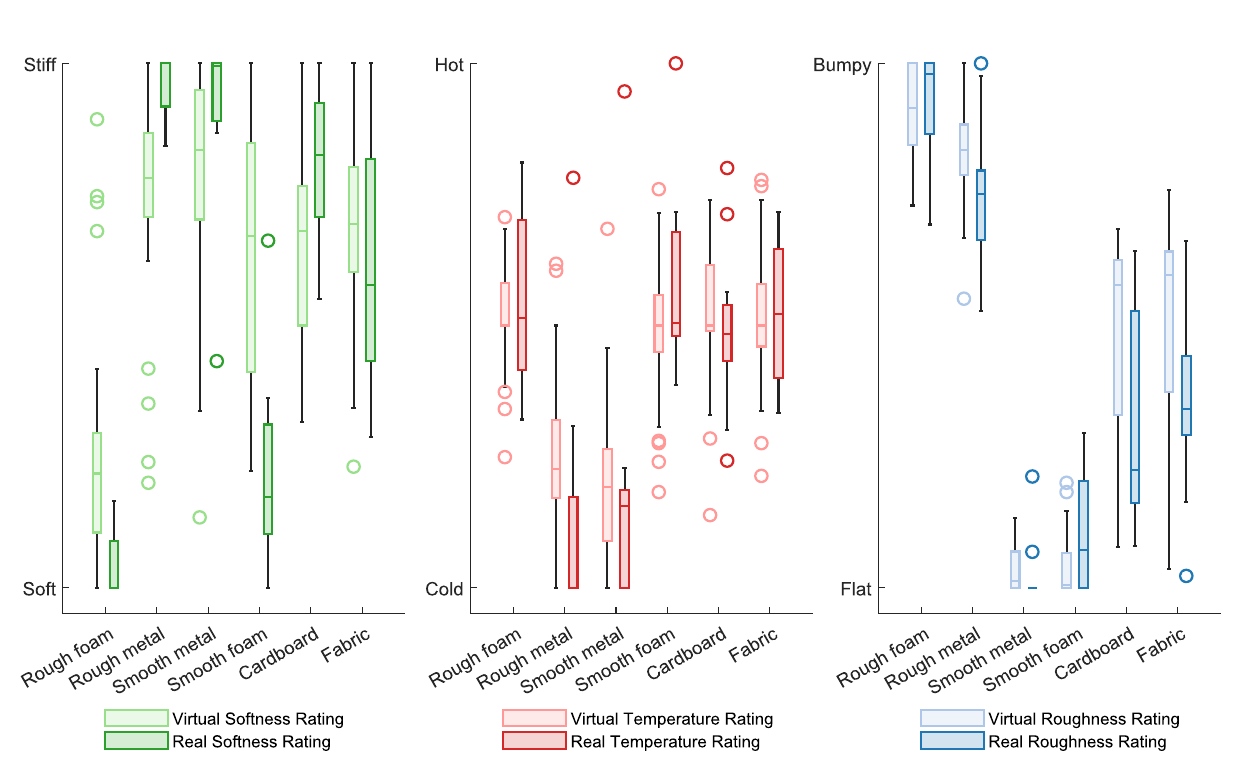}
\vspace{-0.5cm}
    \caption{Comparison between adjective ratings for the soft ring-displayed virtual textures (light) and the real textures explored with guided actions (dark), for their roughness, temperature, and softness dimensions.}
    \label{fig:AdjectiveRatingComparison}
\end{figure*}


\label{app4}
\section{Thermal Control Logic and Actuation Flow}

The thermal control logic is inspired by the approach in \cite{Choi2018}, who employed a PID controller to regulate a Peltier-based thermal display. In contrast, due to the slow thermal dynamics of our water-based system, we implemented only a proportional controller that controls the hydraulic pumps directing water from the hot and cold tanks to the mixing tank, given by

\begin{equation}\label{eq:prop_controller}
    r(t) = K_p \cdot e(t) = K_p \cdot (T_t - T_c)
\end{equation}

where the proportional gain $K_p$ equals 17.5 and $T_t$ and $T_c$ represent the target and current display temperatures, respectively, both normalized to a temperature range between 45 and $10^{\circ}$~C. Equation~\ref{eq:prop_controller} gives a control signal $r(t)$, which after normalization to the $[-1, 1]$ range determines the PWM parameters driving the pumps. Positive $r(t)$ values triggered heating, while negative values triggered cooling. As the measured temperature approaches the target, the PWM values decrease progressively.

\begin{figure}
    \centering
    \includegraphics[width=0.8\linewidth]{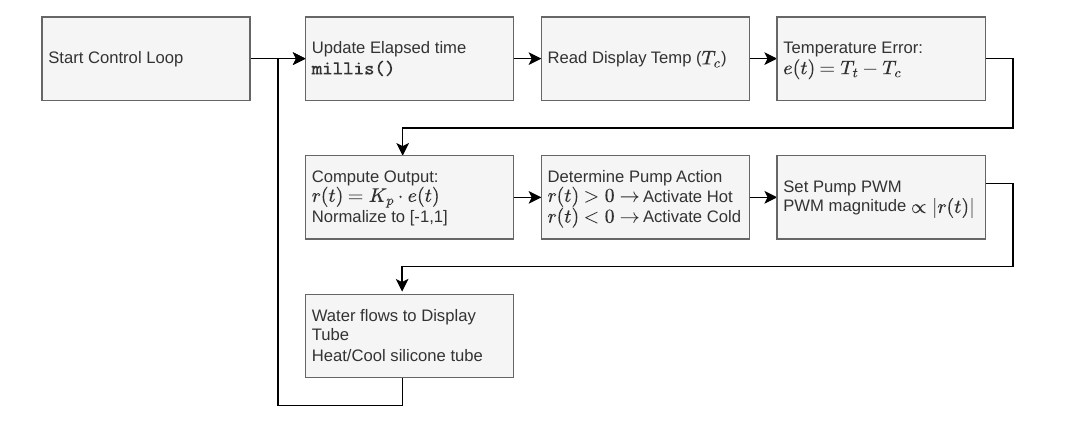}
    \caption{Flowchart of the proportional control loop for regulating the display temperature showing the actuation sequence for a predefined target temperature $T_t$. }
    \label{fig:control}
\end{figure}

The overall control sequence, executed by the microcontroller at 500~ms intervals, is summarized in Figure~\ref{fig:control}

\label{app:thermal-control-logic}

\newpage
\bibliographystyle{elsarticle-harv} 
\bibliography{references}

\end{document}